\documentclass[fleqn,usenatbib]{mnras}

\usepackage{newtxtext,newtxmath}

\usepackage[T1]{fontenc}
\usepackage{ae,aecompl}

\usepackage{graphicx}	
\usepackage{amsmath}	
\usepackage{amssymb}	
\usepackage{longtable}
\usepackage{appendix}
\usepackage{soul}

\usepackage{cprotect}

\newcommand{\gtc}{{\it GTC }}
\newcommand{\osiris}{{\it OSIRIS }}

\usepackage[usenames, dvipsnames]{xcolor}
\usepackage{multirow}
\usepackage{lastpage}

\title[\gtc \osiris DR1]{The {\it Gran Telescopio Canarias} \osiris Broad Band First Data Release}

\author[Cort\'es-Contreras et al.]{
M. Cort\'es-Contreras,$^{1,2}$\thanks{E-mail: mcortes@cab.inta-csic.es}
H. Bouy,$^{3}$
E. Solano,$^{1,2}$
M. Mahlke,$^{1,2}$
F. Jim\'enez-Esteban,$^{1,2}$
\newauthor
J.M. Alacid$^{1,2}$
and C. Rodrigo$^{1,2}$
\\
$^{1}$Departmento de Astrof\'{\i}sica, Centro de Astrobiolog\'{\i}a (CSIC-INTA), ESAC Campus, Camino Bajo del Castillo s/n\\
E-28692 Villanueva de la Ca\~nada, Madrid, Spain\\
$^{2}$Spanish Virtual Observatory, Spain\\
$^{3}$ Laboratoire d'astrophysique de Bordeaux. Universit\'{e} de Bordeaux. CNRS, B18N All\'{e}e Geoffroy Saint-Hilaire, 33615 Pessac, France.
}

\date{Accepted XXX. Received YYY; in original form ZZZ}

\pubyear{2019}

\begin{document}
\label{firstpage}
\pagerange{\pageref{firstpage}--\pageref{lastpage}}
\maketitle

\begin{abstract}
We present the first release of \gtc \osiris Broad Band data archive. This is an effort conducted in the framework of the Spanish Virtual Observatory to help optimize science from the Gran Telescopio Canarias Archive. Data Release 1 includes 6\,788 broad-band images in the Sloan $griz$ filters obtained between April 2009 and January 2014 and the associated catalogue with roughly 6.23 million detections of more than 630\,000 unique sources.
The catalogue contains standard PSF and Kron aperture photometry with a mean accuracy better than 0.09 and 0.15\,mag, respectively. The relative astrometric residuals are always better than 30\,mas and better than 15\,mas in most cases. The absolute astrometric uncertainty of the catalogue is of 0.12\,arcsec.
In this paper we describe the procedure followed to build the image archive and the associated catalogue, as well as the quality tests carried out for validation. To illustrate some of the scientific potential of the catalogue, we also provide two examples of its scientific exploitation: discovery and identification of asteroids and cool dwarfs.
\end{abstract}



\section{Introduction}

Astronomical data archives and catalogues have become a new paradigm in the astrophysics research. Reduced (photometrically and astrometrically corrected images, spectra ready for immediate scientific exploitation,...) and high-level (catalogues, mosaics, stacked images,...) data products are of fundamental importance for archives as they enhance their use by the community. Moreover, these science-ready data products provide a higher visibility of the project results as clearly demonstrated by the large number of refereed papers based on archived data from projects like SDSS \citep{York00}, 2MASS \citep{2003yCat.2246....0C}, UKIDSS \citep{Lawrence07} and WISE \citep{Wright10}, to name a few. By using these resources, astronomers are able to conduct research projects that would otherwise be very time-consuming or completely unaffordable.

The Gran Telescopio de Canarias\footnote{\url{http://www.gtc.iac.es/}} (\gtc), with its 36 individual hexagonal segments acting as a 10.4\,m diameter single mirror, is currently the largest optical-infrared telescope in the world. Operated by the Instituto de Astrof\'isica de Canarias, it is located at the Observatorio del Roque de Los Muchachos in La Palma (Spain), which provides excellent observing conditions.

\begin{figure}
\centering
\includegraphics[width=0.45\textwidth]{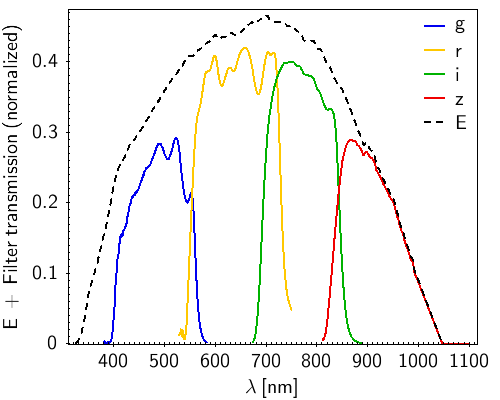}
\caption{Transmission curves of the Sloan g' (blue), r' (yellow), i' (green) and z' (red) filters and the \osiris detection efficiency curve (dashed black).}
\label{fig:curvesfilt}
\end{figure}

\osiris (Optical System for Imaging and low-Intermediate-Resolution Integrated Spectroscopy) is an imager and spectrograph for the optical wavelength range, located in the Nasmyth-B focus of {\it GTC}. \osiris allows broadband imaging over a field of view of 7.8 $\times$ 7.8 arcmin unvignetted covering a spectral range from $\lambda$\,3\,650\,\AA \, to $\lambda$\,10\,000\,\AA, with a high transmission coefficient, in particular at longer wavelengths. The spectral range is covered by the Sloan system broadband filters: u' ($\lambda$\,3\,500\,\AA), g' ($\lambda$\,4\,750\,\AA), r' ($\lambda$\,6\,300\,\AA), i' ($\lambda$\,7\,800\,\AA), z' ($\lambda$\,9\,250\,\AA).
The filter tranmission and detector efficiency curves, taken from the Filter Profile Service maintained by the Spanish Virtual Observatory\footnote{\url{http://svo2.cab.inta-csic.es/theory/fps/}}, are shown in Figure~\ref{fig:curvesfilt}.
More information on the \osiris capabilities can be found at the \gtc web page\footnote{\url{http://www.gtc.iac.es/instruments/osiris/}}.

The \gtc archive\footnote{\url{http://gtc.sdc.cab.inta-csic.es/}} is in operation since 2011 and is hosted by the Spanish Virtual Observatory (SVO\footnote{\url{http://svo.cab.inta-csic.es/main/index.php}}), one of the 21 national and trans-national Virtual Observatory initiatives distributed worldwide and coordinated by the International Virtual Observatory Alliance\footnote{\url{http://ivoa.net}}.

This  paper  presents  the first release of the \osiris Broad Band data. We begin in Sections~\ref{sec:data_curation} and \ref{sec:data_processing} with a description of the data curation, preparation and processing of the images, followed by the source extraction in Section~\ref{sec:sources_extraction}. We present astrometric and photometric calibrations in Sections~\ref{sec:astrometric_calibration} and ~\ref{sec:flux_calibration}. We then turn, in Section~\ref{sec:catalogue_sources}, to a description of the catalogue sources and a photometric validation exercise in which Pan-Starrs DR1 data are compared to \osiris observations. In Section~\ref{sec:science}, we introduce two science cases carried out with the catalogue as an attempt to show its potential for science exploitation. The paper ends with a description on how to access the data and a summary in Sections~\ref{sec:data_access} and \ref{sec:conclusions}, respectively.

\section{Data curation and preparation}\label{sec:data_curation}

The first step consisted in the data selection and curation. Large facilities such as \gtc produce vast amount of data coming from different instruments and observing modes. Since the goal of the present project is to deliver science grade products of the broad-band images, we therefore rejected spectroscopic observations and tunable filter observations taken with \osiris. 

We obtain 27\,470 raw broad-band images from the \gtc Public Archive acquired between April 2009 and January 2014 in the $griz$ bands.
A total of 5\,638 among them were obtained using windowing (moving target images) and were discarded for the present release. The vast majority of them were targeting a few transiting targets requiring a high temporal frequency, and the loss in spatial coverage is minimum.
Thus, it is important to remark that the absence of a source in the catalogue does not imply the absence of \gtc \osiris observations in that position of the sky. To check this point, the \gtc Archive must be used.

A significant number (13\,230) of early images were missing the standard \verb|MJD-OBS| keyword giving the precise acquisition date, which is crucial for many studies focusing on the time-domain.
Although it is possible to recover the missing information through other non-standard FITS keywords (e.g. \verb|OPENTIME|), we will focus in this first data release on the the 8\,602 (27\,470 - 5\,638 - 13\,230) images with the standard \verb|MJD-OBS| keyword.
Among them, there are 858 calibration images that will not be included in the catalogue. Hence, we process 7\,744 \osiris images in the $griz$ filters.

\section{Data processing}\label{sec:data_processing}

The individual raw images were processed using an upgraded version of \emph{Alambic} \citep{2002SPIE.4847..123V}, a software suite developed and optimized for the processing of large multi-CCD imagers, which was adapted for \osiris. 
\emph{Alambic} includes standard processing procedures such as overscan, bias and dark subtraction for each individual readout ports of each CCD, flat-field correction, bad pixel identification and  masking, CCD-to-CCD gain harmonization, and fringing correction in the $z$-band. \emph{Alambic} also combines the flat-field and bad pixel mask into a weight map for each individual dataset, and estimates the sky background in each image using an iterative  multi-resolution median filtering. The method is very efficient, fast and robust for most stellar or extragalactic fields, but will produce artefacts for input images including large extended sources, as well as around the halo of very bright stars. Images and catalogues including large extended sources (i.e. covering a significant fraction of the field-of-view) should therefore be considered with caution, and probably reprocessed from scratch by the user. 

Flat-fields were computed using the best twilight flat-field frames obtained over time windows of 15 days, since the \osiris flat-fields are considered to be very stable over periods of several weeks. 

\section{Source extraction}\label{sec:sources_extraction}

Sources brighter than the 3-$\sigma $ noise of the local background were detected and their photometry and position were measured using {\sc SExtractor} \citep{1996A&AS..117..393B} and {\sc PSFEx} \citep{2013ascl.soft01001B}. The weight-maps were used to properly modulate the detection threshold over the image. Three fixed apertures were used:
\begin{itemize}
\item 21, 31 and 41 pixels in 1$\times$1 binning mode, corresponding to 2.66, 3.94 and 5.21\,arcsec, respectively
\item 11, 15 and 21 pixels in 2$\times$2 binning mode, corresponding to 2.79, 3.81 and 5.33\,arcsec, respectively.
\end{itemize}

In addition to these three fixed apertures, the Kron or automatic aperture as by {\sc SExtractor} \verb|MAG_AUTO|, the PSF (\verb|MAG_PSF|) and the model (\verb|MAG_MODEL|) photometry were also measured. A PSF model was fitted to the data with {\sc PSFEX} using a second order polynomial to model the PSF variations across the field-of-view. The PSF model was then used to fit every source using:
\begin{itemize}
\item standard PSF-fitting
\item a two-dimensional S\'ersic model convolved with the PSF model
\end{itemize}
The {\sc SExtractor} model (i.e., PSF + S\'ersic) parameters offer the advantage of being suited for both point-like and extended sources. Most galaxies are indeed resolved under sub-arcsecond seeing, and the PSF will not give a good fit.

Saturation was set at 62\,000 counts, even though the detector is expected to behave linearly up to the 16 bits encoding limit. 
This is an approximate value and, therefore, some saturated sources may still be present in the catalogue of extracted sources (see Section~\ref{sec:removal}). For a better identification of saturated pixels we will use {\sc MaxiMask} \citep{maximask} in future data releases.

\begin{figure}
\centering
\includegraphics[width=0.45\textwidth]{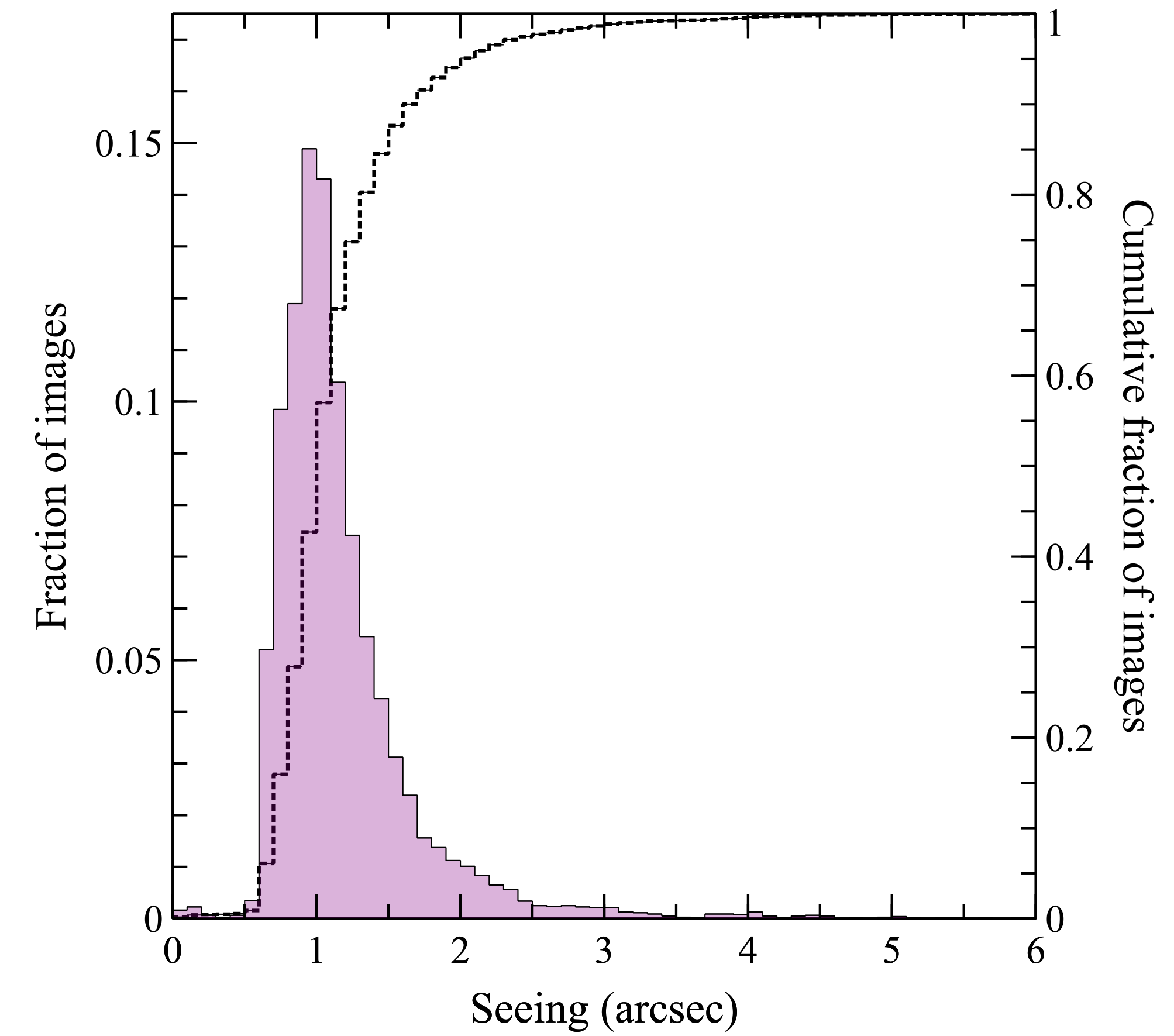}
\caption{Distribution of average FWHM measured in the images themselves by {\sc PSFEx}.
\label{fig:seeing}}
\end{figure}

A few morphometric parameters were also extracted. {\sc SExtractor} offers the possibility to measure the Full-width at half maximum (FWHM), the flux radius (defined as the circular aperture radius enclosing half the total flux), the elongation and the ellipticity. The model fitting also provides useful information about the morphometry and is particularly interesting for extended and saturated sources \citep{2013A&A...554A.101B}.
Figure~\ref{fig:seeing} shows the distribution of FWHM as measured by {\sc PSFEx}. The distribution peaks around 1\,arcsec and more than half of the images were obtained under sub-arcsecond seeing. Note that the distribution extends far towards large values thanks to the \gtc filler program specifically designed to make use of poor ambient conditions.

The {\sc SExtractor} and {\sc PSFEx} configuration files used to produce the catalogues, as well as the parameters obtained are given in Tables~\ref{tab.def_sex}, \ref{tab.def_psfex} and \ref{tab.def_params}.

\section{Astrometric calibration}\label{sec:astrometric_calibration}

\begin{figure*}
\centering
\includegraphics[width=0.9\textwidth]{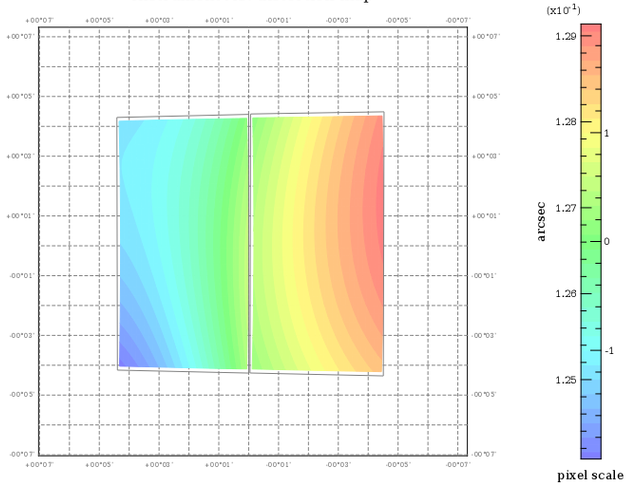}
\caption{Distortion map for the $i$-band filter as measured by {\sc Scamp}. \label{fig:distortions}}
\end{figure*}

A precise astrometric calibration was computed using {\sc Scamp} \citep{2006ASPC..351..112B}. The 2MASS catalogue was used as reference for most frames, except when too few 2MASS sources were available in the field-of-view to derive a proper astrometric solution. In those cases, the SDSS (DR10) \citep{2014ApJS..211...17A} or USNO-B1 \citep{Monet03} catalogues were used.
The absolute astrometric accuracy is therefore set by these catalogues and is expected to be better than 0.1\,arcsec. Non-linear geometric distortions were fitted using a third order polynomial, as illustrated in Fig.~\ref{fig:distortions}. The internal accuracy is estimated to be better than 30\,mas in all cases, and better than 15\,mas in most cases.
Celestial coordinates are given at the observing epoch and Equinox J2000.0.

\begin{figure*}
\centering
\includegraphics[width=0.95\textwidth]{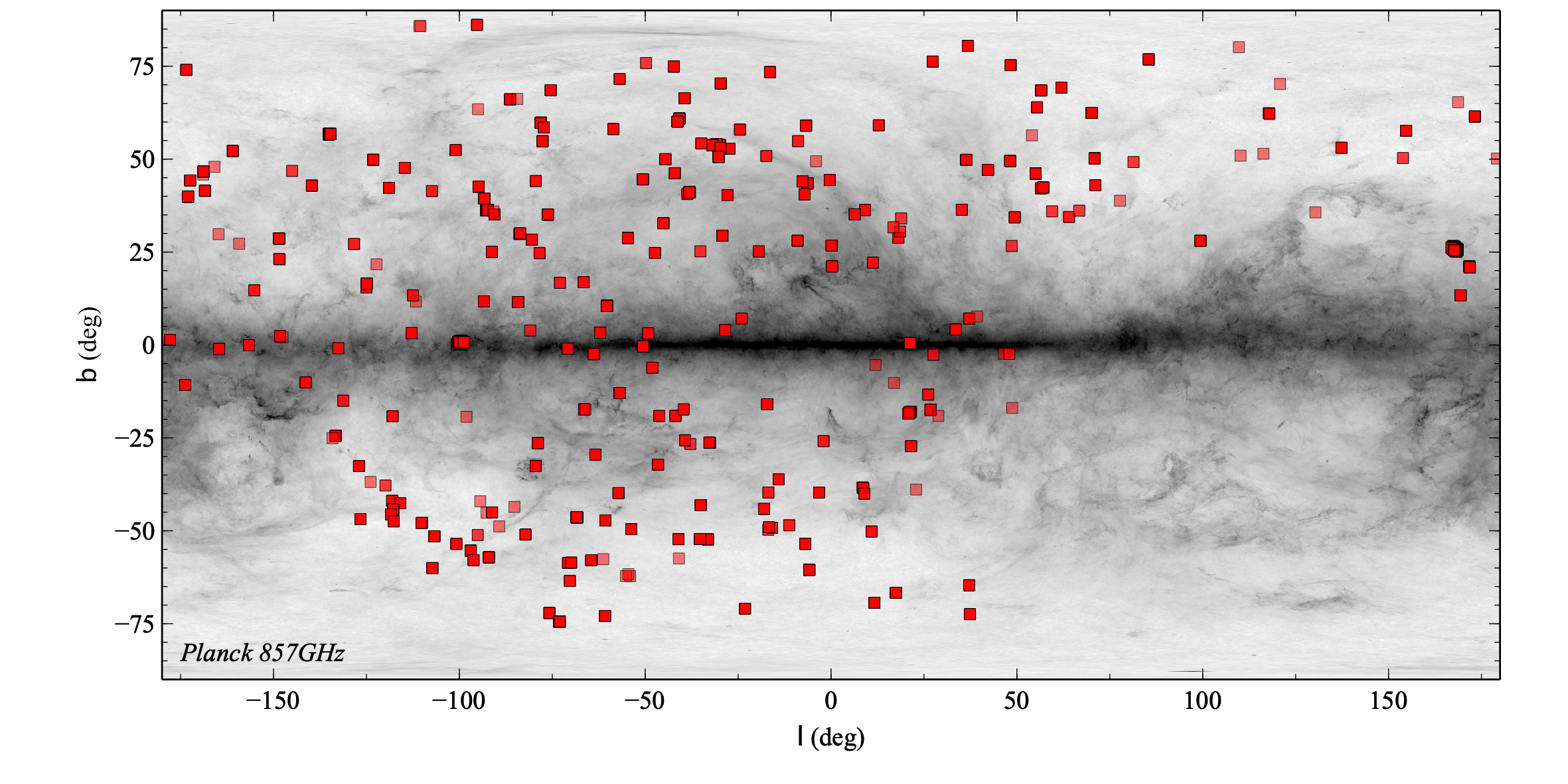}
\caption{Location of the 7\,214 \osiris pointings overplotted on an all-sky 857GHz {\it Planck} map.}
\label{fig:coverage}
\end{figure*}

The astrometric calibration allowed us to detect 530 problematic images, typically including technical images with no sources at all, images obtained during a tracking failure or having suffered a read-out failure. The total number of remaining images added to 7\,214, distributed over the entire northern sky accessible from La Palma, as illustrated in Fig.~\ref{fig:coverage}.

This figure reflects the patchy nature of the \osiris observations, with a modest fraction (0.02\%) of the full sky being covered (8.05\,deg$^{2}$). This is a primary difference with typical survey projects covering a regular geometric pattern in the sky and with uniform properties, such as exposure time and filter set.

Individual astrometrically calibrated images can be easily downloaded from the \gtc archive and stacked to make a larger and/or deeper mosaic of a given field. Because the astrometric calibration was obtained using the Astr{\it O}matic software suite, we recommend to use {\sc SWarp} \citep{2002ASPC..281..228B} for the stacking. {\sc Swarp} will indeed properly interpret the astrometric solution included in the image headers and the weight map delivered with every individual image by simply using the "\verb|-WEIGHT_TYPE MAP_WEIGHT|" option. \\

\section{Flux calibration} \label{sec:flux_calibration}

The set of \osiris images used to build the catalogue constitute, by definition, a heterogeneous collection where the variety of observational parameters (e.g. exposure times) reflect the diversity of science cases carried out with them. While images with the shortest exposure times typically present bright sources and few or no faint objects, images with the largest exposure times, aimed at reaching fainter sources, usually include saturated objects. This wide range of exposure times reflects both the potential and the challenge of the associated catalogue.

We adopted the Pan-STARRS DR1 survey (PS1 hereafter) \citep{Kaiser10,Chambers16} as the reference to photometrically calibrate \gtc \osiris images.
The PS1 catalogue includes PSF and Kron photometric measurements, which we will use to calibrate our \verb|MAG_PSF| and \verb|MAG_AUTO| photometry, respectively.

In order to obtain the calibration parameters, we searched for counterparts to each \osiris source in PS1 within 1.0\,arcsec. If more than one counterpart exists in the search radius we took the nearest one.
Each individual CCD subimage has been calibrated independently.
For a source to be considered as a calibration source, it must fulfill the following requirements:
\begin{itemize}

\item PS1 sources must be point-like sources with \verb|PSF|-\verb|Kron|<0.05\,mag \citep{Chambers16}, and {\it Qual} flag equal to 52.

\item PS1 sources must be within the saturation and detection limits of the survey as stated in \cite{Chambers16}.

\item Magnitude errors, both in PS1 and \osiris, must be smaller than 0.2\,mag.

\item In order to avoid spikes and extended sources, we select in our catalogue sources whose FWHM, elongation and ellipticity do not significantly deviate from the median value of all sources in the subimage.

\end{itemize}

With the selected sources, we carried out a sigma-clipping linear fit in each subimage, after which we performed the calibration only if there were more than six calibrating stars satisfying the criteria listed above. Besides, we provide calibrated magnitudes only if the Pearson correlation coefficient ($r$) of the linear fit is greater than 0.98.
Of the 14\,428 subimages available (7\,214$\times$2), we calibrated 13\,555 and 13\,403 in \verb|MAG_PSF| and \verb|MAG_AUTO| photometry, respectively. Of them, 12\,454 and 11\,614 have a correlation coefficient $r \geq 0.98$ (82\% and 87\%, respectively). A total number of 11\,045 subimages have both, PSF and automatic photometric calibrations with $r \geq 0.98$.
The differences in the number of calibrated subimages arise, on one hand, from the discrepancies in magnitude errors in PSF and AUTO photometry, which prevent the same detection to pass the magnitude cuts in both photometries. On the other hand, the different procedures of flux measurements by {\sc SExtractor} in few cases lead to a wrong measure of the \verb|FLUX_PSF| but not of \verb|FLUX_AUTO| for the same source and viceversa. We could therefore obtain a varying number of sources in the same subimage for PSF and AUTO calibrations.
The fact that the \osiris observations lie within the Pan-STARRS footprint ensures that, in all cases, the lack of photometric calibration is not due to a different spatial coverage of the two surveys but to the absence of sources fulfilling the criteria listed above or to the bad quality of the fit.

\begin{figure*}
\centering
\includegraphics[width=0.246\textwidth]{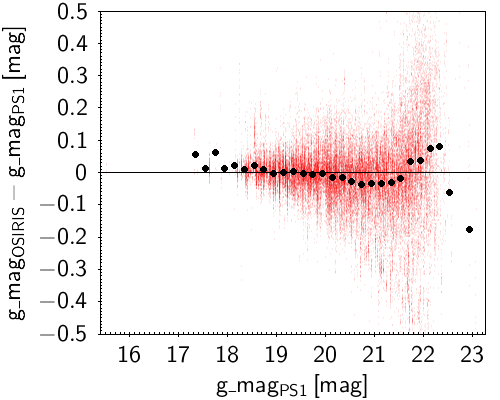}
\includegraphics[width=0.246\textwidth]{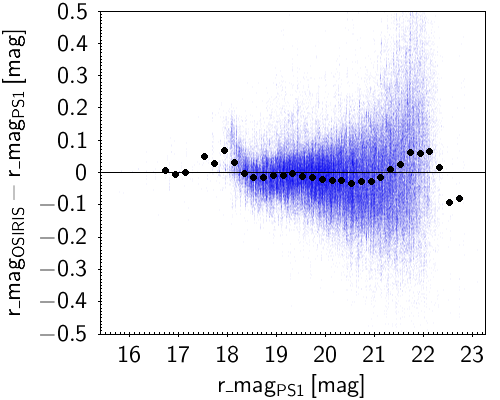}
\includegraphics[width=0.246\textwidth]{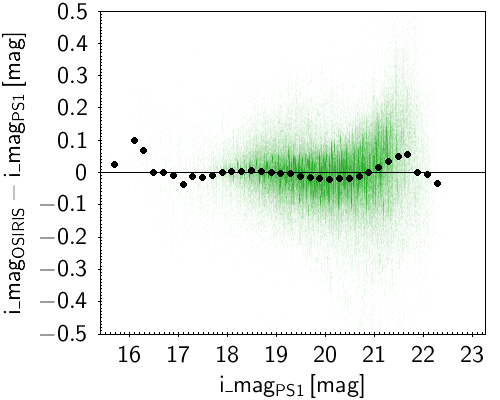}
\includegraphics[width=0.246\textwidth]{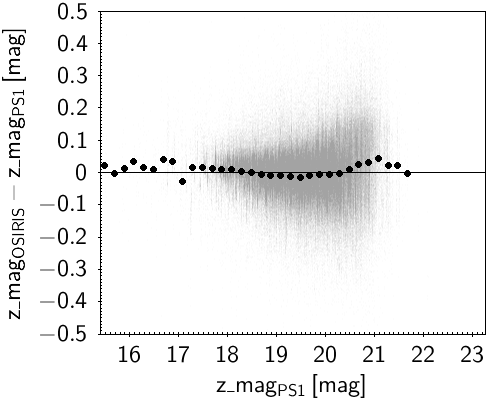}
\includegraphics[width=0.246\textwidth]{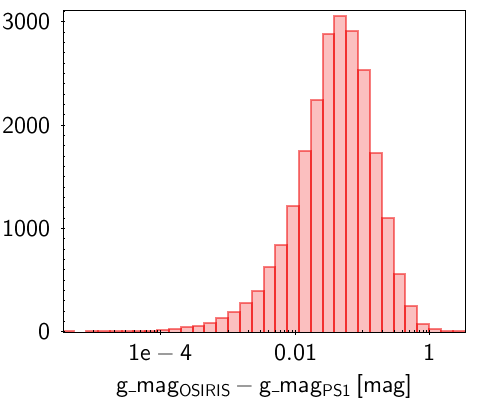}
\includegraphics[width=0.246\textwidth]{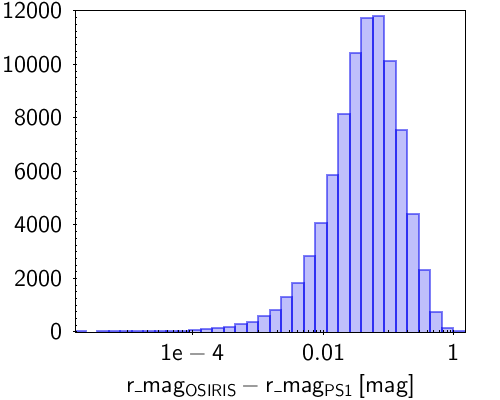}
\includegraphics[width=0.246\textwidth]{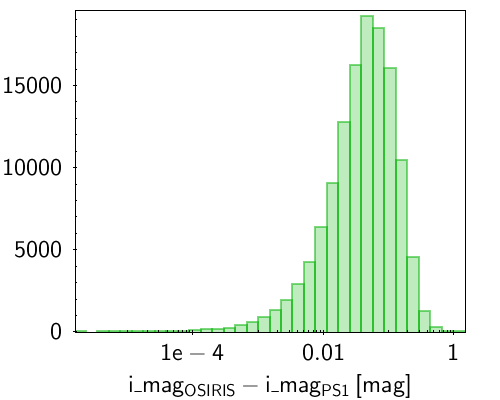}
\includegraphics[width=0.246\textwidth]{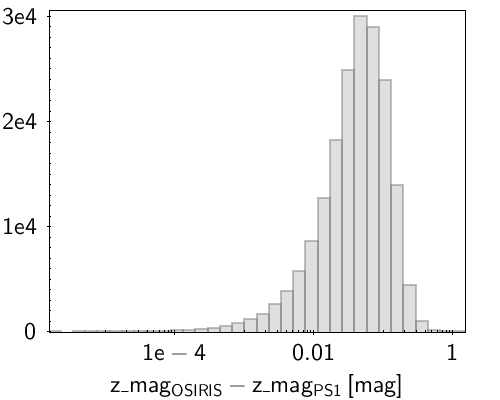}
\includegraphics[width=0.246\textwidth]{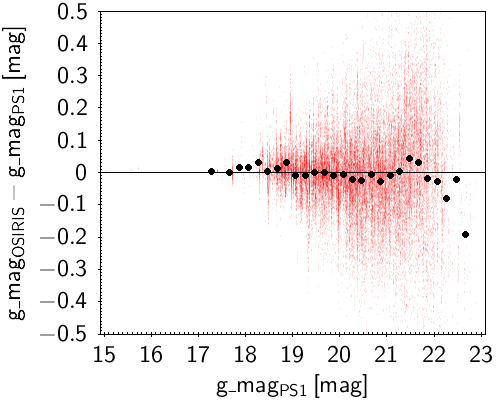}
\includegraphics[width=0.246\textwidth]{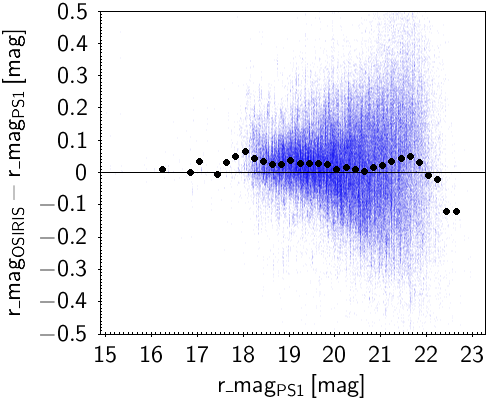}
\includegraphics[width=0.246\textwidth]{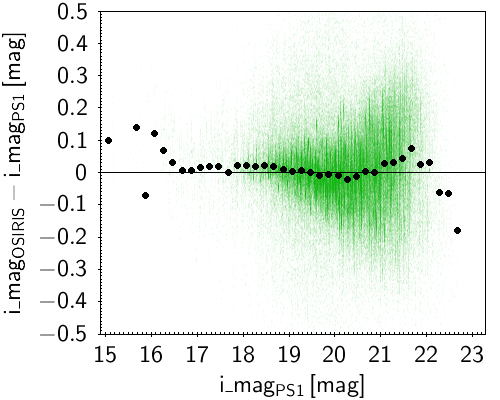}
\includegraphics[width=0.246\textwidth]{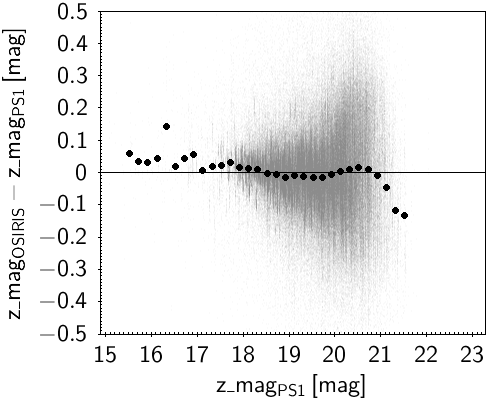}
\includegraphics[width=0.246\textwidth]{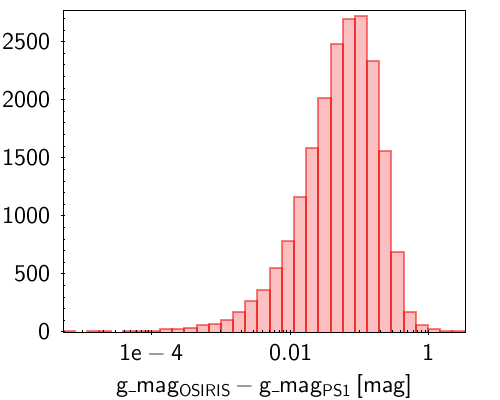}
\includegraphics[width=0.246\textwidth]{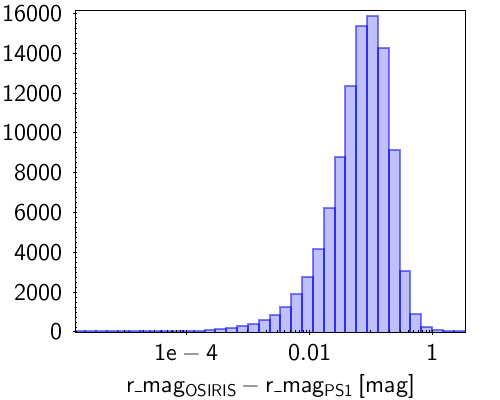}
\includegraphics[width=0.246\textwidth]{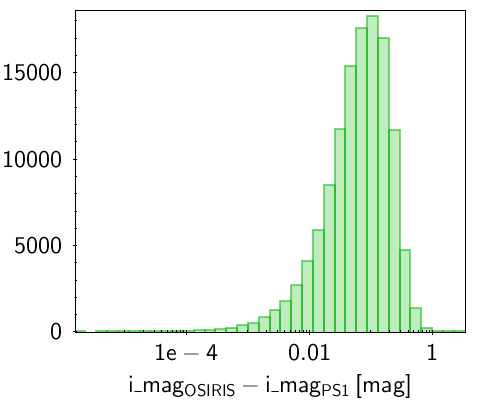}
\includegraphics[width=0.246\textwidth]{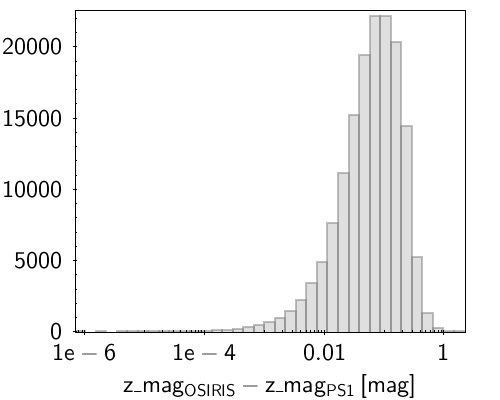}
\caption{Comparison between PS1 and \osiris magnitudes of the sources used for the PSF (top first and second panels) and AUTO (third and fourth panels) photometric calibrations with $r >= 0.98$. Black filled circles represent the average difference of magnitudes in bins of 0.2\,mag. Only bins with more than ten sources are represented.
\label{fig:cal_comparison}}
\end{figure*}

Figure~\ref{fig:cal_comparison} shows the differences between PS1 and \osiris calibrated magnitudes of the calibration sources as a function of PS1 magnitudes and the distributions of the absolute values of the magnitude differences taken from the 12\,454 and 11\,614 subimages with good linear fit in PSF (top first and second panels) and AUTO (third and fourth panels) photometry, respectively. The few number of outliers existent in the first and the third panels, typically associated to sources with larger magnitude errors that pass the criteria listed above and that remain after the linear fit within $3\sigma$, have a minor impact in the calibration as we are using a weighted linear fit where the weight is inversely proportional to the error in magnitude.
We show in the first and third panels the average magnitude differences in bins of 0.2\,mag. Only bins with more than ten points are shown.
In all bands, these values are larger for faint sources (over $\sim$21.5\,mag in $g$, $r$ and $i$ bands, and over $\sim$20.0\,mag in the $z$ band). We also observe an increasing trend towards the bright edge of the plots. This behaviour is observed in magnitude intervals with less than 100 sources, very few compared to the several thousands included at intermediate magnitude bins.
These averaged differences of magnitudes remain under 0.1\,mag in PSF and under 0.2\,mag in AUTO comparisons for all bands.
On average, AUTO photometry presents larger scatter compared with PSF. 
Mean magnitude absolute differences (second and fourth panels) are marginal. They vary from 0.063 to 0.079\,mag in PSF photometry and from 0.096 to 0.098\,mag in AUTO photometry, depending on the filter.

\begin{figure*}
\centering
\includegraphics[width=0.4\textwidth]{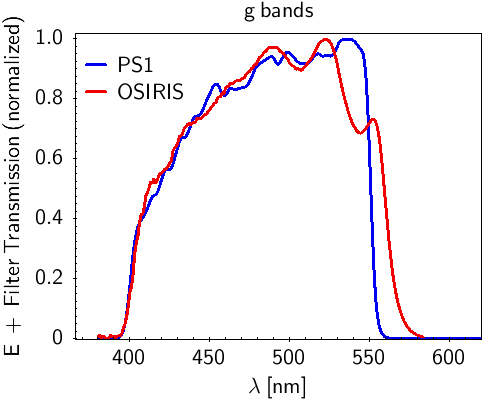}
\includegraphics[width=0.4\textwidth]{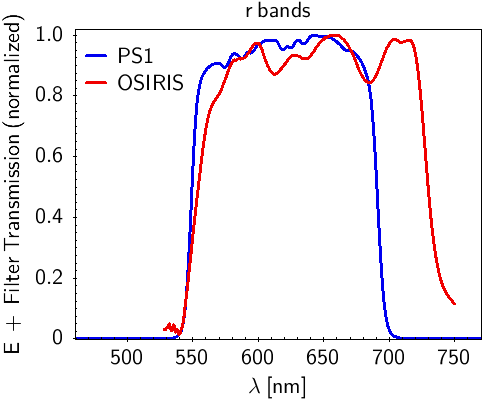}
\includegraphics[width=0.4\textwidth]{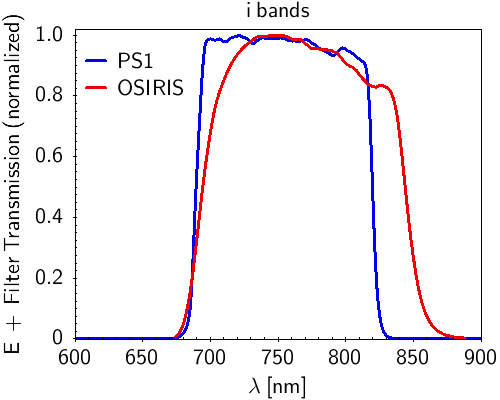}
\includegraphics[width=0.4\textwidth]{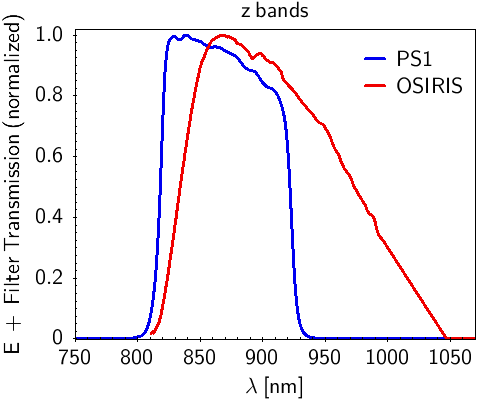}
\caption{Comparison between the transmission curves of \osiris and PS1 in each band.
\label{fig:tcurves_comp}}
\end{figure*}

The difference in the filter transmission curves of the PS1 and \osiris filters do also contribute to the discrepancy between magnitudes. Figure~\ref{fig:tcurves_comp} shows the comparison between the transmission curves for each filter. In both cases, the throughput of the instrument is considered. In the case of the PS1 curves, taken from the SVO Filter Profile Service, we removed the contribution of the atmosphere, for comparison with \osiris' curves. While the $g$-band filters match fairly well, \osiris includes redder wavelengths in the $r$, $i$ and $z$ bands.
To assess these differences in the calibrated photometry, we address a colour dependence test in Section~\ref{sec.colour}.

The number of calibration sources per subimage ranges from 7 to 1\,510 for \verb|MAG_PSF| and from 7 to 1\,323 for \verb|MAG_AUTO|, with a mean number of 68 and 65 stars, and a standard deviation of 113 and 107, respectively.

The contribution to the total error budget of the calibration errors is, on average, between 0.05\,mag and 0.07\,mag in \verb|MAG_PSF| and between 0.09\,mag and 0.12\,mag for $z$ in \verb|MAG_AUTO|, depending on the filter.

\begin{figure}
\centering
\includegraphics[width=0.45\textwidth]{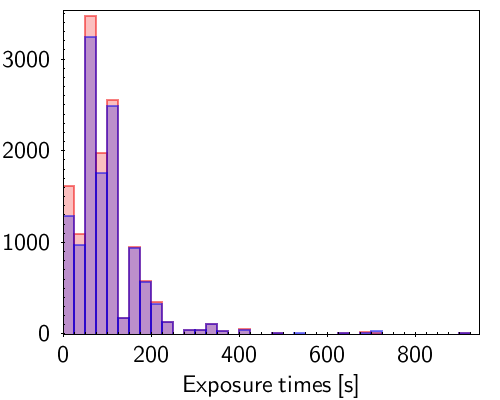}
\cprotect\caption{Histogram of the exposure times of \verb|MAG_PSF| (red) and \verb|MAG_AUTO| (blue) calibrated subimages with $r >= 0.98$.}
\label{fig:exptimes_cal}
\end{figure}

Figure~\ref{fig:exptimes_cal} shows the exposure time distributions of the calibrated subimages in \verb|MAG_PSF| and \verb|MAG_AUTO| photometry. In the whole catalogue, exposure time ranges from 0.5 to 900\,s.
From the mean value of the distributions ($\sim$ 100\,s) we can estimate the mean limiting magnitudes of the catalogue at 24.4, 23.9, 23.3, 22.1\,mag at S/N=5, and 23.6, 23.1, 22.6, 21.4\,mag at S/N=10 for $g$, $r$, $i$ and $z$ respectively. However, for the longest exposures, these magnitudes can reach 25.8, 24.1, 23.6 and 22.4 magnitudes for $griz$ at S/N=5. 

To assess the photometric quality of our calibrations, we selected all calibration sources with magnitudes in the range  $(18.5,18.5,18.5,18.0) < ($g$,$r$,$i$,$z$) < (22.0,22.0,22.0,21.5)$ and with at least five detections with PSF and AUTO photometry, separately. These intervals of magnitudes were chosen in order to get a representative sample of the sources used for the calibration (over 90\% for each filter and photometry), and to avoid saturation and very faint sources. We obtained the standard deviations of the measured magnitudes for each of the repeated sources. The median of the standard deviations in each filter are $(\sigma_g,\sigma_r,\sigma_i,\sigma_z)=(0.034,0.026,0.038,0.034)$ magnitudes for PSF and $(\sigma_g,\sigma_r,\sigma_i,\sigma_z)=(0.043,0.038,0.053,0.056)$ magnitudes for AUTO photometry. These low values reflect the good performance of the calibration parameters over time despite the assorted character of the observations.

We summarize in Table~\ref{tab.calibration_prop} the detailed numbers of the associated contribution of the calibration to the photometric errors, the range of exposure time, the mean limiting magnitudes at SNR=5 and SNR=10, and the typical magnitude deviations for the repeated sources in the above defined sample, for each filter.

\begin{table*}
        \centering
        \caption {Summary of the photometric properties of the calibrated subimages.}
        \label{tab.calibration_prop}
        \begin{tabular}{l cccc cccc }
        \hline \hline
        \noalign{\smallskip}
    &   \multicolumn{4}{c}{PSF} &   \multicolumn{4}{c}{AUTO}    \\
    &   $g$ &   $r$ &   $i$ &   $z$ &   $g$ &   $r$ &   $i$ &   $z$    \\
        \noalign{\smallskip}
        \hline
        \noalign{\smallskip}
        \noalign{\smallskip}

Contribution to the     &   &   &   &   &   &   &   &   \\
photometric error (mag)   &   0.06 &   0.07    &   0.05    &   0.05    &  0.12    &   0.012   &   0.09    &   0.10    \\
\noalign{\smallskip}
Exposure times (s)  &   1--900  &   1--360  &   0.5--360    &   0.5--300 &   1-900  &   1--542  &   0.5--360    &   0.5--30 \\
\noalign{\smallskip}
Mean limiting magnitudes    &   &   &   &   &   &   &   &   \\
at SNR=5 (mag)    &   24.4 &   23.9    &   23.3    &   22.1    &   &   &   &   \\
\noalign{\smallskip}
Mean limiting magnitudes    &   &   &   &   &   &   &   &   \\
at SNR=10 (mag)   &   23.6 &   23.1    &   22.6    &   21.4    &   &   &   &   \\
\noalign{\smallskip}
Limiting magnitudes for the     &   &   &   &   &   &   &   &   \\
longest exposures and SNR=5 (mag)    &   25.8    &   24.1    &   23.6    &   22.4  &   &   &   &   \\
\noalign{\smallskip}
\hline
\noalign{\smallskip}
    &   \multicolumn{8}{c}{Photometric quality} \\
\noalign{\smallskip}
Magnitude interval (mag)  &    18.5--22.0 &   18.5--22.0  &   18.5--22.0  &   18.0--21.5  &     18.5--22.0 &   18.5--22.0  &   18.5--22.0  &   18.0--21.5              \\
Median of the std. dev. $\sigma$ (mag)    &  0.034    &   0.026   &   0.038   &   0.034   &   0.043 &   0.038   &   0.053   &   0.056   \\ 
        \noalign{\smallskip}    
        \hline
        \end{tabular}
\end{table*}

\section{Catalogue sources}\label{sec:catalogue_sources}

\subsection{Construction}

\subsubsection{Removal of spurious and saturated detections}\label{sec:removal}

To build the cleanest possible catalogue, we first removed sources with PSF instrumental magnitude errors equal to zero, larger than 1\,mag, or equal to 99 (meaning that {\sc SExtractor} PSF fit did not converge). We also discarded sources with the {\sc SExtractor} keywords \verb|FLUX_MAX|, \verb|FLUX_RADIUS| (defined as the half-light radius in Section~\ref{sec:sources_extraction}) and \verb|FWHM_IMAGE| smaller than or equal to zero, \verb|FLAGS_WEIGHT| (weighted extraction flag related to the presence of close neighbours bright enough to significantly bias the photometry, bad pixels, blended objects, saturated pixels or other features) equal to two (note that we do not impose any condition on the extraction flags \verb|FLAGS| parameter), and \verb|SNR_WIN| (the window-based signal-to-noise ratio) smaller than five or equal to 1e30 (the latter, related to a bad extraction of the source).
The limit in \verb|SNR_WIN| at five is set to avoid sources with very poor photometry.

\begin{figure}
\centering
\includegraphics[width=0.45\textwidth]{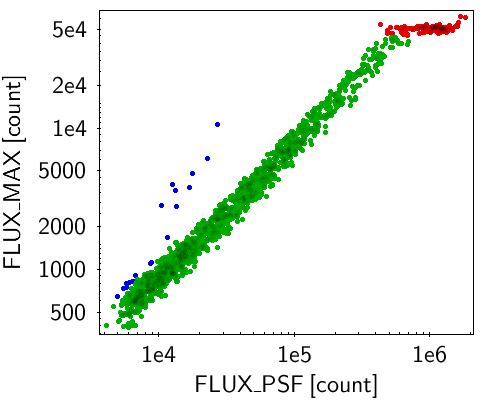}
\cprotect\caption{Maximum flux at the peak (\verb|FLUX_MAX|) versus \verb|FLUX_PSF| in logarithmic scale for detections in one subimage. Blue points represent spurious detections, red points represent saturated sources and green points stand for valid detections.}
\label{fig:sat}
\end{figure}

Although a saturation limit was set at 62\,000 counts when extracting the sources from the images, few saturated sources remain. Therefore, for each subimage in the catalogue, we removed saturated sources as well as bad pixels, cosmic rays and artifacts by accounting for the linear relation between the flux at the peak of the distribution (\verb|FLUX_MAX|) and the integrated flux (\verb|FLUX_PSF|), as shown in Figure~\ref{fig:sat}. The procedure to remove spurious sources on each subimage consisted of running an iterative process that discards first all detections which ratio \verb|FLUX_MAX|/\verb|FLUX_PSF| deviates by more than 2$\sigma$ from the mean value. With the mean and standard deviation values of the remaining detections, we define a critical value of  \verb|FLUX_MAX|/\verb|FLUX_PSF| equal to the mean value plus 3$\sigma$. From the original catalogue of sources of the subimage, we removed all detections which flux ratio is greater than this critical value. The rejected detections are represented with blue points in Figure~\ref{fig:sat}.

Once the removal of spurious sources has been completed, we removed saturated sources by identifying the position in the relation \verb|FLUX_MAX| vs. \verb|FLUX_PSF| at which the detector breaks linearity.
To determine this position, we iteratively performed a linear fit starting with sources with the lowest \verb|FLUX_MAX| and increasing towards higher values of \verb|FLUX_MAX|. We consider that linearity starts to fail when the Pearson correlation coefficient $r$ is lower than 0.98 and keeps decreasing. When there were no data available, we decreased the correlation coefficient to 0.96. We then selected the upper half of the sample with highest \verb|FLUX_MAX| to avoid non-linear behaviour at low fluxes (i.e., faint sources) and followed two complementary approaches:

\begin{itemize}

\item On one hand, we took the maximum number of sources which linear fit provides a correlation coefficient greater than 0.98 (or 0.96 if it is the case). Since this condition itself does not ensure the removal of all saturated sources, we defined a subset with the detections above that maximum number (i.e., sources with higher flux) and established a cut in \verb|FLUX_MAX| defined as the minimum value of the \verb|FLUX_MAX| in the subset minus its standard deviation.

\item On the other hand, we looked for the position at which the slope of the fit starts to decrease and defined a subset with detections starting at that point and with increasing fluxes. We again established a cut  in \verb|FLUX_MAX| defined as the minimum value of the \verb|FLUX_MAX| in the subset minus its standard deviation. When the standard deviation was smaller than 2\,000 counts, we subtracted twice the value of the standard deviation to ensure the removal of all controversial sources. 

\end{itemize}
 
In both cases, we removed all detections with \verb|FLUX_MAX| above those limits.

\begin{figure}
\centering
\includegraphics[width=0.45\textwidth]{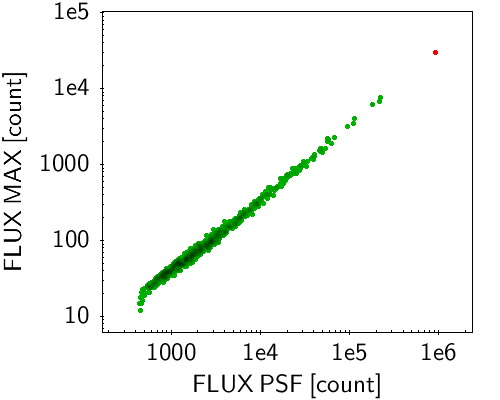}
\cprotect\caption{Maximum flux at the peak (\verb|FLUX_MAX|) versus \verb|FLUX_PSF| in logarithmic scale for detections in one subimage after the removal of spurious and saturated detections following the first two steps. Green points represent valid non-saturated source detections and the red point stands for a saturated source that follows the linear behaviour between fluxes and that accounts with more than 9\,000\,counts in \verb|FLUX_MAX| with respect to the maximum flux of the brightest-non saturated source in the subimage.}
\label{fig:sat9000}
\end{figure}

Despite the completion of these steps, we still found a few saturated sources that happen to follow the linear relation between fluxes as shown in Figure~\ref{fig:sat9000} and, therefore, have not been automatically removed. We observed that saturated sources have fluxes (\verb|FLUX_MAX|) more than $\approx$9\,000\,counts higher than the corresponding flux to the brightest-non saturated source in the subimage. Hence, we removed detections which \verb|FLUX_MAX| is at least 9\,000\,counts greater than the brightest, non-saturated source in the subimage.

The above described procedure does only relate to PSF photometry. Hence, in parallel to this, we built a second catalogue considering exclusively calibrated \verb|MAG_AUTO| magnitudes and following similar steps than before. The only difference comes from the identification of saturated sources, because the relation between \verb|FLUX_MAX| and \verb|FLUX_AUTO| displays a higher dispersion and, therefore, executing the procedure to remove saturated sources as described above prevented us from obtaining a complete catalogue in AUTO photometry, since sources in more than 50\% of the calibrated subimages did not fulfill the required criteria. Therefore, for each subimage, we took the cuts in \verb|FLUX_MAX| determined when removing saturated sources from PSF photometry and eliminated all sources in this second catalogue with \verb|FLUX_MAX| higher than those values. This approach limits the AUTO photometry to the 11\,045 subimages that have both, PSF and AUTO calibrated photometry.

Moreover, we included a photometric flag to each detection in the catalogue to account for magnitudes beyond the magnitude coverage of the PS1 sources used in the photometric calibration in PSF (\verb|Flag_psf|) and AUTO (\verb|Flag_auto|) photometry. "B" and "C" stand for magnitudes fainter and brighter than the magnitude coverage of PS1, respectively. Detections with magnitudes in between are flagged with "A".

\subsubsection{Morphologic classification}\label{subsec:morph}

In a second step, we evaluated in our catalogue the magnitude difference limit \verb|PSF| $-$ \verb|AUTO| at 0.05\,mag proposed in the PS1 catalogue \citep{Farrow2014} to separate point-like (\verb|PSF| $-$ \verb|AUTO| $<0.05$\,mag) and extended sources (\verb|PSF| $-$ \verb|AUTO| $>0.05$\,mag).
Of the \osiris detections in the $g$-band under 21\,mag (limit above which the high dispersion makes this classification unreliable) and over $\sim$17\,mag (magnitude limit imposed by PS1) with PS1 counterparts and that satisfy PS1 \verb|gmag| $-$ \verb|gKmag| $<0.05$\,mag (i.e., point-like sources), 95.5\% were also classified as point-like sources following the same criterion with \osiris photometry (i.e., had \verb|PSF| $-$ \verb|AUTO| $<0.05$\,mag).
Hence, we could consider this photometric criterion to be valid to select point-like sources. However and due to its magnitude restriction, we do not ascribe an extended or point-like source flag in the catalogue following this rule. We evaluated instead the separation between point-like and extended sources at any magnitude range by looking into the \verb|FWHM_IMAGE|, \verb|elongation| and \verb|ellipticity| parameters.
To do so, we selected sources in each subimage with PSF calibrated magnitudes between 17 and 21\,mag (regardless the filter) and with magnitude differences \verb|PSF| $-$ \verb|AUTO|<0.05\,mag, and compute their \verb|FWHM_IMAGE|, \verb|elongation| and \verb|ellipticity| mean values and standard deviations.
Sources in the subimages which parameter values differed by more than the mean value within 2$\sigma$ were tagged as extended sources and as point-like sources otherwise.

To assess the goodness of this criterion, we compared both, the \osiris photometric and morphometric classifications, in the interval between 17 and 21\,mag. In all bands, more than 92\% of the sources identified photometrically as point-like were also labeled as point-like using morphometric parameters. Moreover, extended sources were also in agreement in more than 75, 83, 74, and 72 per cent of the cases in the $g$, $r$, $i$, and $z$ bands, respectively. We therefore conclude that the morphometric criterion can be extended to all detections in the catalogue, regardless their magnitudes, to tabulate them as point-like or extended sources.

For 1\,378 subimages in the four bands, we could not apply this selection because of the lack of automatic (Kron) magnitudes due to the absence of photometric calibration, or the lack of sources with magnitude differences under 0.05\,mag and magnitudes between 17 and 21\,mag. Since the number of subimages without any classification was significantly large, we decided to use PS1 magnitudes to identify point-like sources and compute their \verb|FWHM_IMAGE|, \verb|elongation| and \verb|ellipticity| mean and standard deviations values to be used as a reference for the classification of all detections in the subimage. In these cases, the classification is noted in the catalogue as "P*" for point-like and "E*" for extended sources.

After this, there are still 77 subimages (39\,246 detections representing 0.6\% of the catalogue) for which we were not able to ascribe any classification.

The privileged location and weather conditions of the Observatory favour astronomical observations with a typical seeing of 1\,arsec and reaching often lower values. Detections with FWHM below 0.5\,arcsec were removed as they were just simply artifacts. Detections with FWHM over 7\,arcsec and/or with ellipticities over 0.7 were also removed. The mean ellipticity measured in point-like sources is 0.11$\pm$0.07 with just a 1.9\% showing an ellipticity worse than {\it e} >0.3. This confirms the good performance of the \gtc tracking capability.

\subsubsection{Resulting catalogue}\label{subsec:resulting_cat}

The next step is to build a science-ready, user-friendly catalogue containing both astrometric and photometric information as well as flags to warm about quality issues. 

In summary, the catalogue contains 6\,226\,520 detections corresponding to 633\,559 different sources in 12\,409 subimages. Detections were merged into sources by performing an internal match of the whole catalogue (this is, regardless the photometric band) within 0.5\,arcsec using STILTS \citep{Taylor06}. This value is a trade-off between completeness and reliability. Larger values may allow unrelated detections to be linked in the same source while smaller values would pose problems for faint sources with large centroiding errors. A detailed description of each column in our detection catalogue is given in Table~\ref{tab.catalogue_description}.

Table~\ref{tab.cat-info} lists, for each filter, the number of subimages and detections contained in the catalogue.

\begin{table}
        \centering
        \caption {Number of subimages and detections per filter in the catalogue.}
        \label{tab.cat-info}
        \begin{tabular}{c c c}
        \hline \hline
        \noalign{\smallskip}
Filter  & Number of &	Number of	  \\
  & subimages         &	detections					\\

        \noalign{\smallskip}
        \hline
        \noalign{\smallskip}
        \noalign{\smallskip}
Sloan g & 1\,328 & 532\,760	 	\\			
Sloan r & 2\,866 & 1\,585\,249		  \\	
  Sloan i & 3\,825 & 2\,041\,184	  \\
Sloan z &  4\,390	&	2\,067\,327	  \\
        \noalign{\smallskip}    
        \hline
        \end{tabular}
\end{table}

Table~\ref{tab.bands-sources} summarizes the number of sources that have been detected in one, two, three or the four bands, regardless which bands are they. Near 30\% of the sources in the catalogue has been detected in two or more bands and only 3.2\% has been detected in the four bands.

\begin{table}
        \centering
        \caption {Number of sources detected in one to four bands.}
        \label{tab.bands-sources}
        \begin{tabular}{c c }
        \hline \hline
        \noalign{\smallskip}
Number of       &   Number of 	  \\
 bands          &   sources		\\

        \noalign{\smallskip}
        \hline
        \noalign{\smallskip}
        \noalign{\smallskip}
4 & 19\,991 (3.2\%)\\			
3 & 64\,350 (10.2\%)\\	
2 & 106\,747 (16.8\%)\\
1 & 442\,471 (69.8\%)\\ 
        \noalign{\smallskip}    
        \hline
        \end{tabular}
\end{table}

Astrometric errors in the catalogue were computed from the quadratic sum of the windowing position errors (i.e., the errors in the ellipse parameters reported by {\sc SExtractor}) and the estimated absolute astrometric calibration uncertainty mentioned in Section~\ref{sec:astrometric_calibration}. The mean accuracy in the catalogue is 0.12\,arcsec.

In order to provide accurate photometry at a single epoch, we identified the best detection for each source, hereafter refereed to as {\it primary} detection. The criteria used to select the best detection were to have signal to noise ratio above the mean value for that source, and the least relative PSF or AUTO magnitude error.
A total of 1\,209\,058 primary detections satisfied the above criteria. Primary detections are flagged in the catalogue as {\it p} while the rest of detections are flagged with {\it s} under {\it Flag\_source}.

In addition, we composed a source catalogue with these {\it primary} detections of each source, for the user to easily access the best photometry. It contains selected data such as the source identifier, equatorial coordinates, PSF and AUTO calibrated magnitudes with their corresponding flags, epochs and url of the associated image in each band, and a source class parameter ({\it cl}) defined as the ratio between the number of detections classified as point-like ("P" or "P*") and the total number of detections. The source class parameter takes values between 0 and 1, being 0 when the source has always been identified as extended and 1 when it has always been identified as point-like.

This source catalogue is complementary to the previously defined one containing all detections, magnitudes and parameters. Table~\ref{tab.primarycatalogue_description} contains a detailed description of each column in the source catalogue.

\begin{figure*}
\centering
\includegraphics[width=0.5\textwidth]{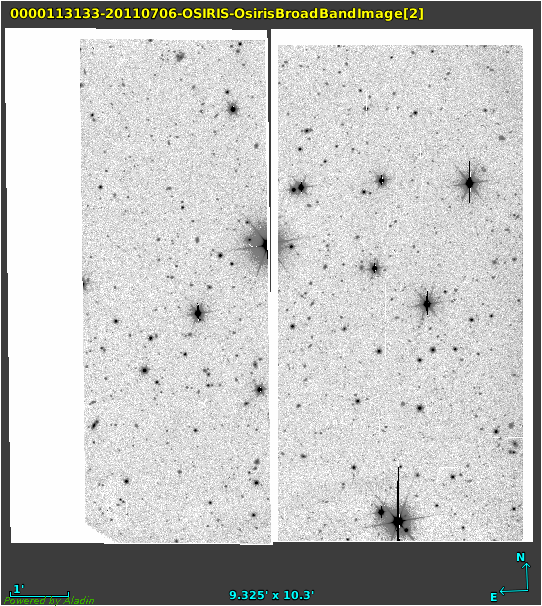}\\
\includegraphics[width=0.48\textwidth]{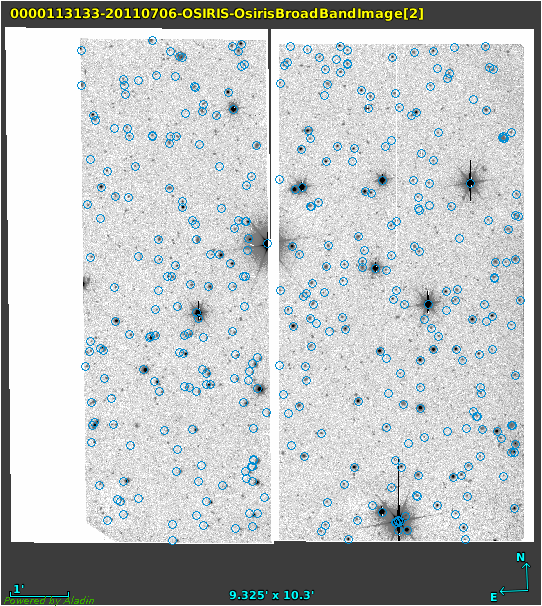}
\includegraphics[width=0.48\textwidth]{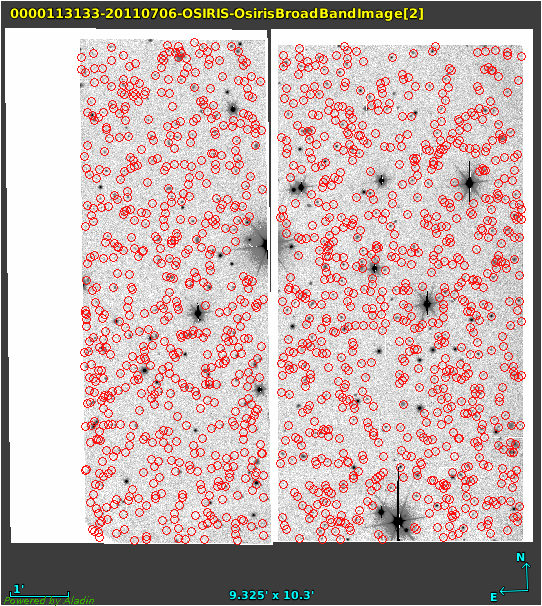}
\caption{Example of an \osiris image centered at 17:06:59.64 +58:46:37.5 (top panel) overlayed with PS1 sources (bottom left panel, blue open circles) and \osiris sources (bottom right panel, red open circles).
\label{fig:campo_gtcps1}}
\end{figure*}

Typical saturation and limiting magnitudes in the catalogue in each filter are 13.1--24.6\,mag in $g$, 13.6--24.5\,mag in $r$, 13.1--23.9\,mag in $i$, and 12.0--22.7\,mag in $z$, respectively. Saturation magnitudes correspond to the minimum value of the PSF magnitude in the catalogue and limiting magnitudes correspond to the 90th percentile, per each filter.
This means that the magnitude limits in the \osiris catalogue are between 1.4 and 2 magnitudes brighter and between 0.4 and 1.4 magnitudes fainter than PS1, depending on the filter. Figure~\ref{fig:campo_gtcps1} shows a \gtc \osiris image together with all PS1 and \gtc sources in the field of view. The increased depth of the \osiris catalogue (1\,378 objects) compared to PS1 (366 objects) is clearly seen. The catalogues have 208 sources in common. There are 158 sources that are in PS1 and not in our catalogue due to saturation or \verb|Flags_weight| parameter equal to two (see the first paragraph in Section~\ref{sec:removal}). Magnitudes of the 1\,170 sources in our catalogue that are not in PS1 range from 20.0 to 25.4\,mag, which depending on the filters, would be beyond the magnitude limit of PS1.

\begin{figure*}
\centering
\includegraphics[width=0.246\textwidth]{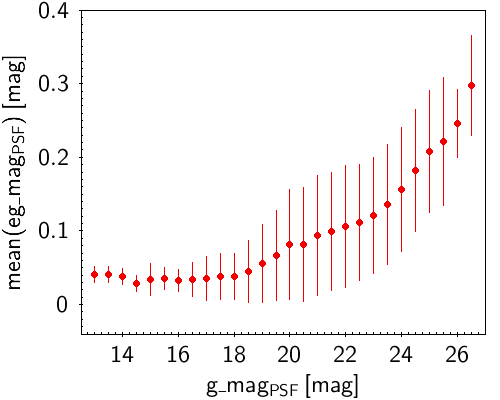}
\includegraphics[width=0.246\textwidth]{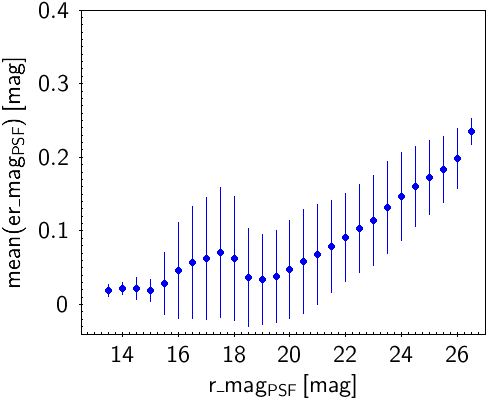}
\includegraphics[width=0.246\textwidth]{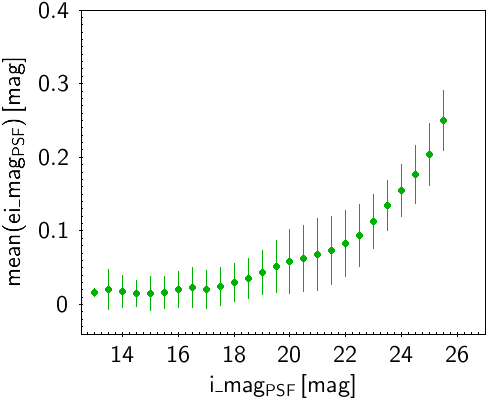}
\includegraphics[width=0.246\textwidth]{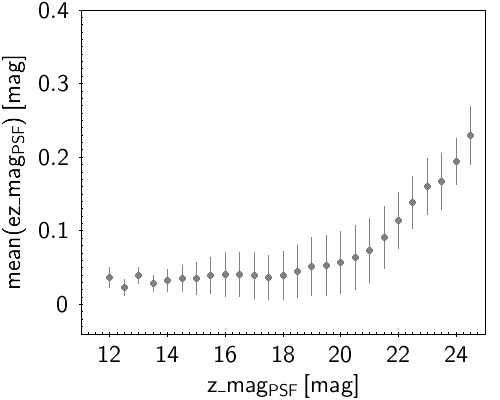}
\includegraphics[width=0.246\textwidth]{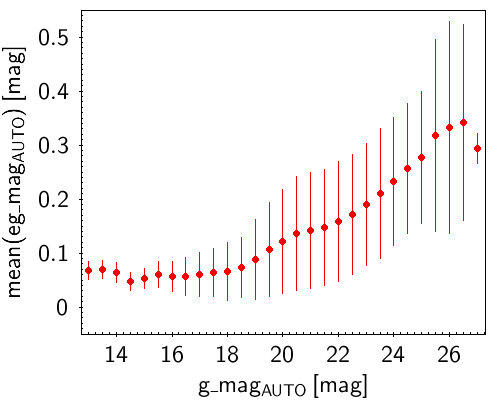} 
\includegraphics[width=0.246\textwidth]{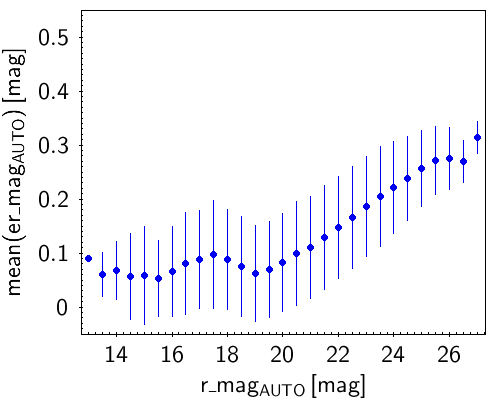}
\includegraphics[width=0.246\textwidth]{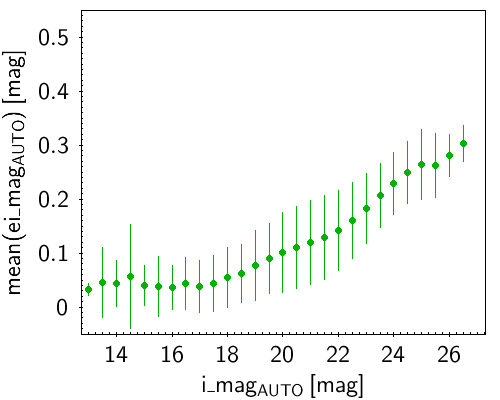}
\includegraphics[width=0.246\textwidth]{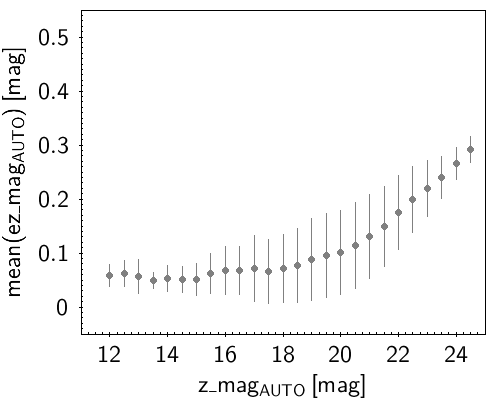}
\cprotect\caption{Photometric errors versus magnitudes for \verb|MAG_PSF| (top) and \verb|MAG_AUTO| (bottom) in the four bands.
Bullets are the average values of the magnitude errors in bin sizes of 0.5\,mag and the error bars show the standard deviation.}
\label{fig:sensitivity}
\end{figure*}

Figure~\ref{fig:sensitivity} represents averaged PSF and AUTO magnitude errors versus magnitude in bins of 0.5\,mag, illustrating the sensitivity reached in the catalogue on each of the four filters. The trend is to increase with increasing magnitudes (i.e., towards fainter sources).
In all bands, mean magnitude errors are under 0.1\,mag up to 21.5 / 19\,mag and do not exceed 0.30 / 0.35\,mag, in PSF /AUTO photometry, respectively. In PSF photometry, the mean photometric accuracy of the catalogue is 0.09\,mag and 0.15\,mag in AUTO photometry.

\subsection {Catalogue quality assessment}

\subsubsection{Comparison with PS1} 

\begin{figure*}
\centering
\includegraphics[width=0.246\textwidth]{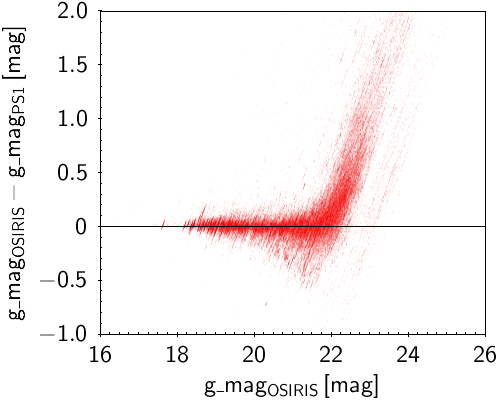}
\includegraphics[width=0.246\textwidth]{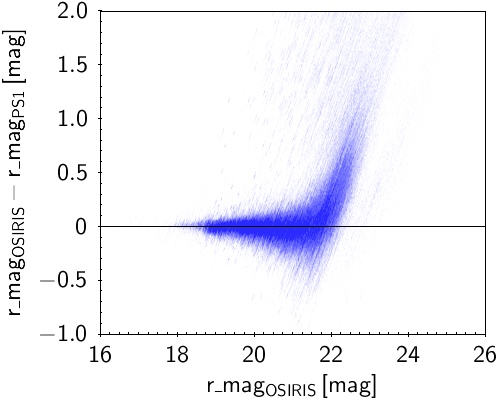}
\includegraphics[width=0.246\textwidth]{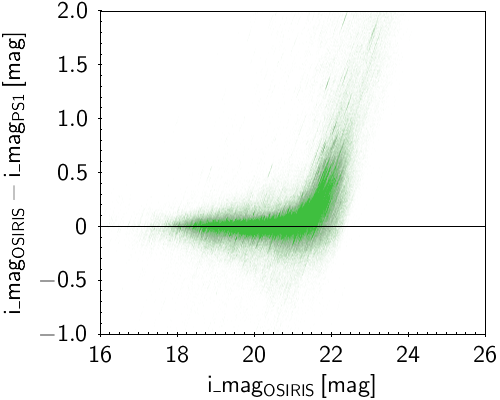}
\includegraphics[width=0.246\textwidth]{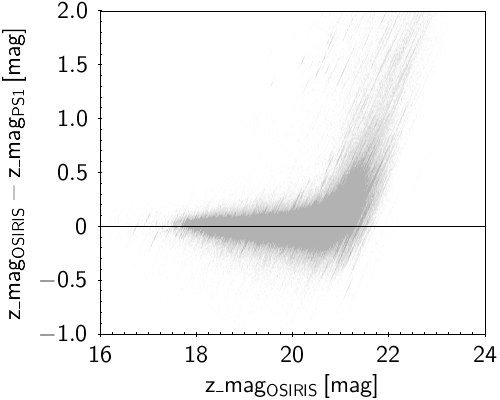}
\includegraphics[width=0.246\textwidth]{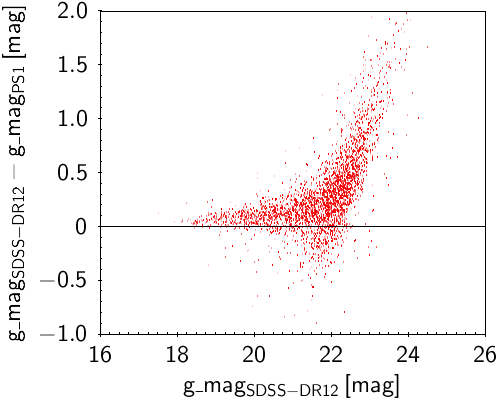}
\includegraphics[width=0.246\textwidth]{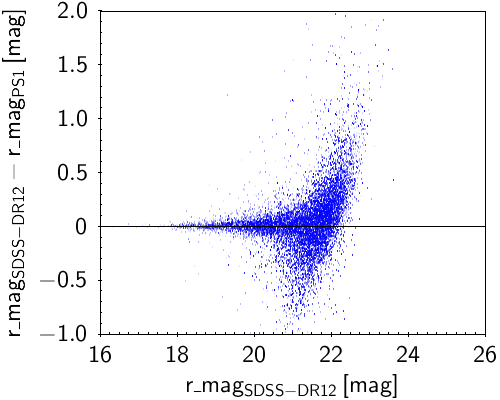}
\includegraphics[width=0.246\textwidth]{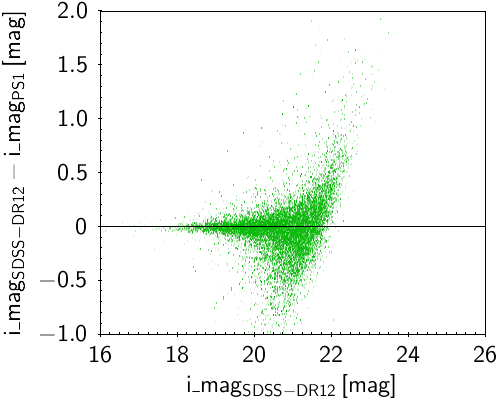}
\includegraphics[width=0.246\textwidth]{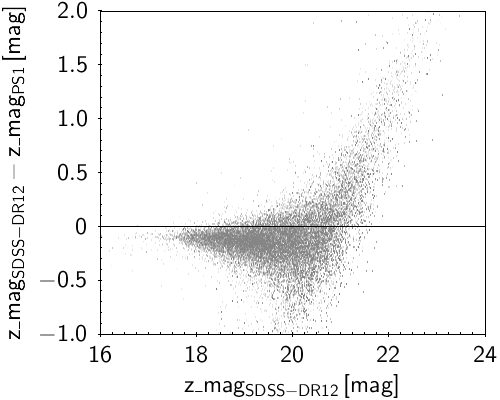}
\caption{Comparison, for each filter, of PSF PS1 magnitudes with PSF \osiris (top panel) and of PSF PS1 magnitudes with PSF SDSS DR12 (bottom panel). 
\label{fig:ps1_vs_gtc}}
\end{figure*}

Top panel in Figure~\ref{fig:ps1_vs_gtc} compares good quality PS1 sources ({\it Qual} equal to 52) versus \osiris PSF magnitudes. We observe good agreement up to PS1 $\sim$21.5\,mag in the $g$ and $r$ bands, $\sim$21.0\,mag in the $i$ band and $\sim$20.5\,mag in the $z$ band, magnitudes above which we observe a tendency towards fainter \osiris magnitudes. Most outliers in these plots do also missmatch when comparing their PS1 magnitudes with clean ({\it q\_mode} equal to '+') SDSS DR12 photometry (see bottom panel in this figure). Note that the number of sources represented in the plots of the bottom panel is lower than the number of sources represented in the plots of the top panel due to the requirement of having good PS1 and SDSS DR12 photometry. In the comparison of PS1 with SDSS DR12 we observe as well the tendency towards fainter magnitudes above 20--21.5\,mag, depending on the filter. Hence, we ascribe these magnitude differences to an intrinsic feature of PS1 photometry.

\begin{figure}
\centering
\includegraphics[width=0.45\textwidth]{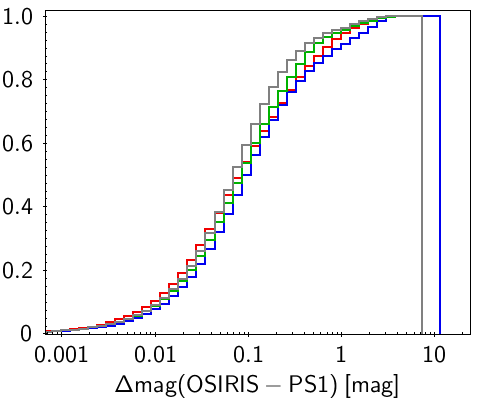}
\caption{Normalized cumulative distribution of PSF absolute magnitude differences between \osiris and PS1 in logarithmic scale in the $g$ (red), $r$ (blue), $i$ (green) and $z$ (gray) bands.
\label{fig:diff_mags}}
\end{figure}

Figure~\ref{fig:diff_mags} shows, for each filter, the normalized cumulative distribution of the PSF magnitude absolute differences between \osiris and PS1 sources with good quality flags ({\it Qual} equal to 52).
For 90\% of the sample of sources in the magnitude interval of linear behaviour, the photometric scatter in absolute values is below 0.16, 0.19, 0.18, and 0.16 magnitudes in $g$, $r$, $i$ and $z$, respectively. This reflects the good photometric agreement between catalogues.

\subsubsection{Colour dependence}\label{sec.colour}

\begin{figure*}
\centering
\includegraphics[width=0.32\textwidth]{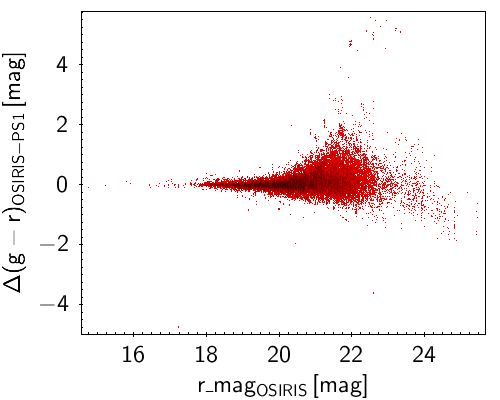}
\includegraphics[width=0.32\textwidth]{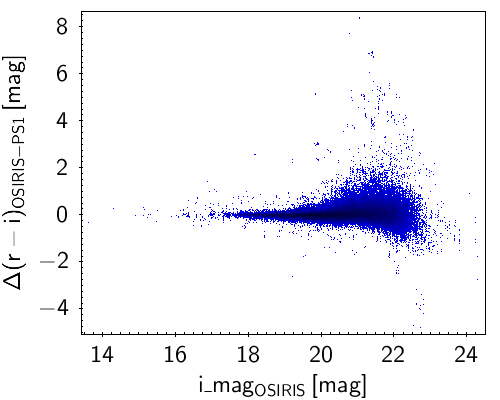}
\includegraphics[width=0.32\textwidth]{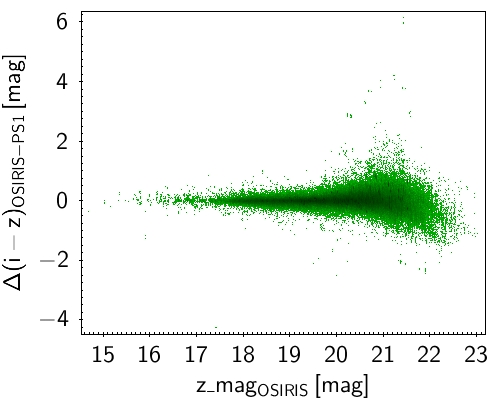}
\includegraphics[width=0.32\textwidth]{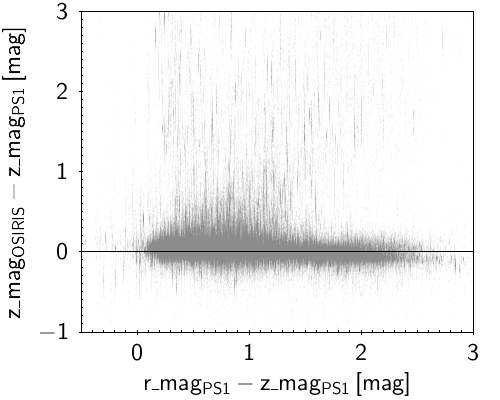}
\includegraphics[width=0.32\textwidth]{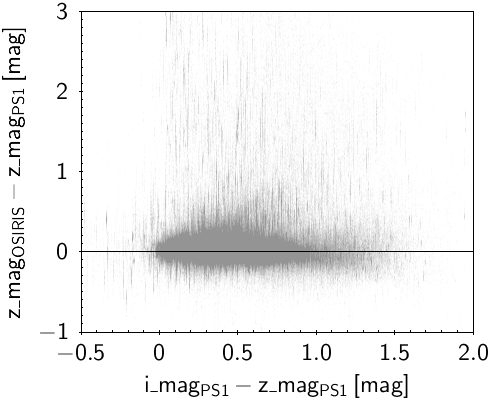}
\caption{Top panel: Dependence of the difference of colours (\osiris -- PS1) with \osiris PSF magnitudes. Bottom panel: Dependence of the difference of magnitudes in the z band with PS1 $r-z$ (left) and $i-z$ (right) colours.
\label{fig:colour_dep}}
\end{figure*}

To investigate the effect of a colour term in the photometric calibration, we compared the colour differences between \osiris and PS1 as a function of the calibrated magnitudes in the top panel of Figure~\ref{fig:colour_dep}. The large dispersion observed towards fainter sources is likely associated to the already noticed magnitude differences over 20--21.5\,mag when comparing \osiris with PS1 (see Figure~\ref{fig:ps1_vs_gtc}). Moreover, the number of sources with significantly high colour differences is not statistically representative (less than 1\% of the plotted sample).
The bottom panel of the figure shows the difference of magnitudes in the z band with respect to the PS1 $r-z$ (left) and $i-z$ colours (right). In light of these plots, we do not observe any colour dependence.

\subsubsection{Binning mode and pixel position dependence}

We verify whether the observed magnitude differences of sources with PSF magnitudes fainter than 20.5--21.5\,mag (depending on the filter) when comparing with PS1 photometry have a relation with the pixel position of the source in the CCD.

\begin{figure*}
\centering
\includegraphics[width=0.246\textwidth]{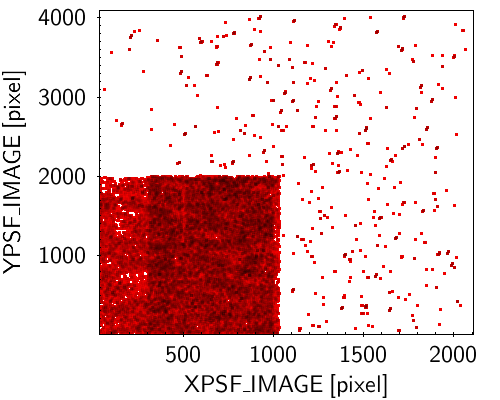}
\includegraphics[width=0.246\textwidth]{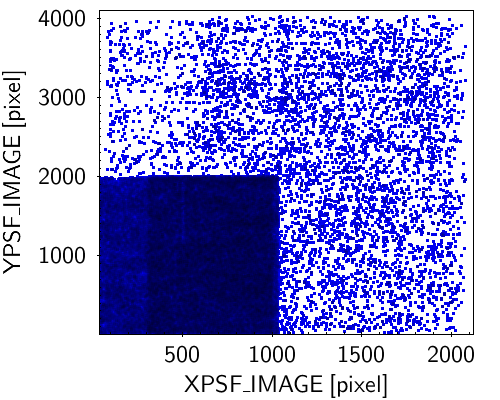}
\includegraphics[width=0.246\textwidth]{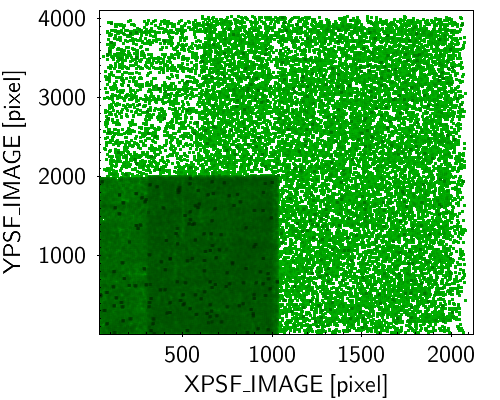}
\includegraphics[width=0.246\textwidth]{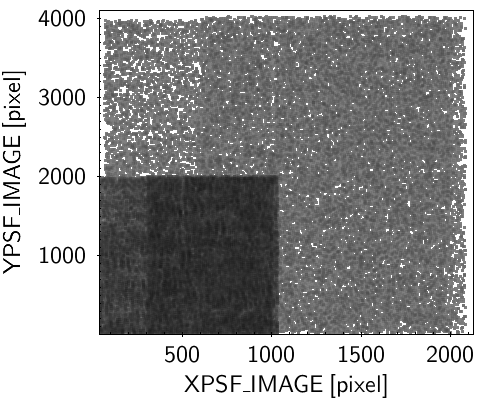}
\caption{CCD $xy$ position of sources with PS1 photometry and fainter PSF magnitudes than 21.5\,mag in $g$ (red panel) and $r$ (blue panel), 21.0\,mag in $i$ (green panel), and 20.5\,mag in $z$ (gray panel).
\label{fig:xy_dep}}
\end{figure*}

Figure~\ref{fig:xy_dep} shows, for each filter, the $xy$ pixel position of sources with PSF magnitudes fainter than 21.5\,mag in $g$ and $r$, 21.0\,mag in $i$, and 20.5\,mag in $z$ (i.e., sources with the largest magnitude deviations) and with good quality PS1 photometry. \osiris standard observing mode uses 2$\times$2 binned pixels, this is, images of $\sim$ 1\,024$\times$2\,048\,pixels size. This is the binning mode where the majority of the observations were performed. In general, binning the pixels aims to increase the signal to noise of measured images at a cost of losing spatial resolution. In light of the random arrangement of the sources in the CCD, we can conclude that there is no dependence of the magnitude differences either with the binning mode of observation or the pixel position.

\subsubsection{Internal photometric precision}

\begin{figure*}
\centering
\includegraphics[width=0.246\textwidth]{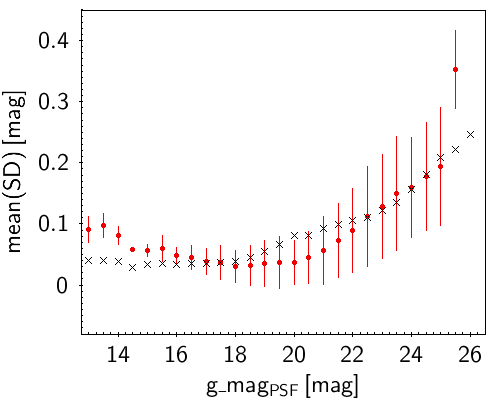}
\includegraphics[width=0.246\textwidth]{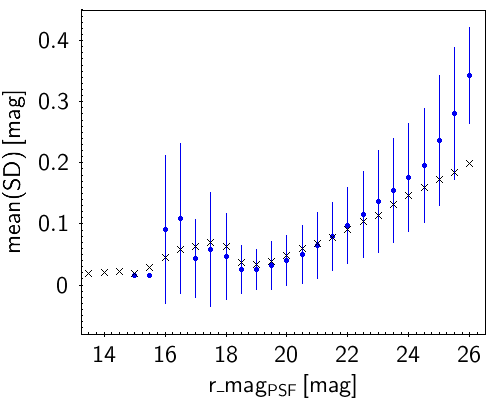}
\includegraphics[width=0.246\textwidth]{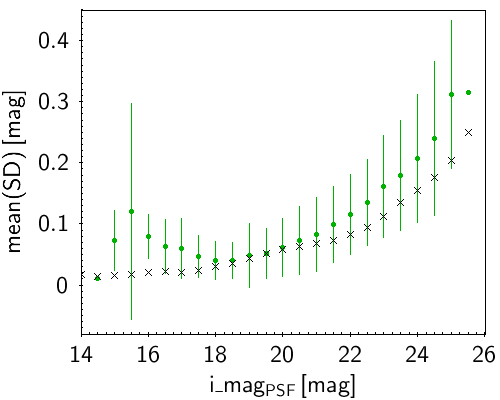}
\includegraphics[width=0.246\textwidth]{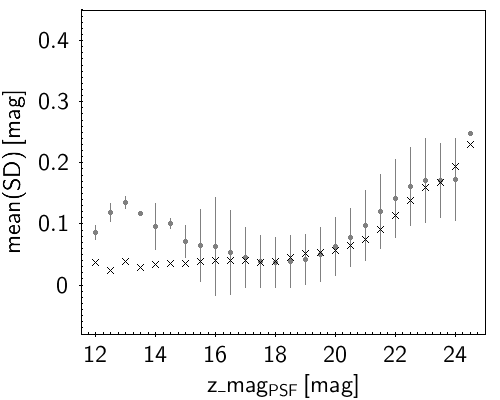}
\includegraphics[width=0.246\textwidth]{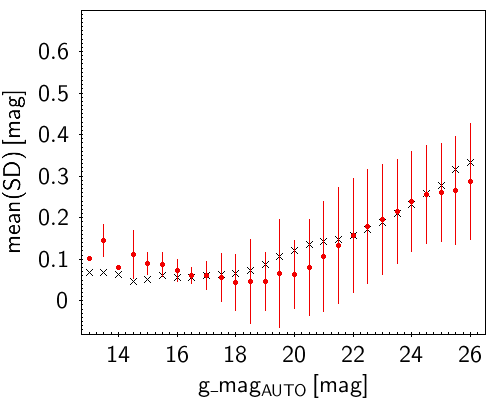}
\includegraphics[width=0.246\textwidth]{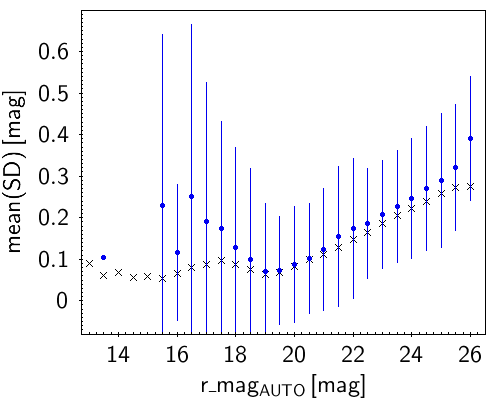}
\includegraphics[width=0.246\textwidth]{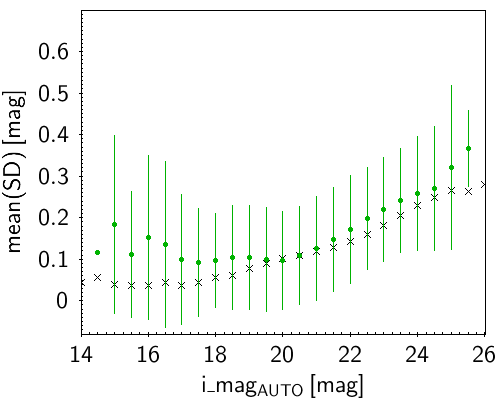}
\includegraphics[width=0.246\textwidth]{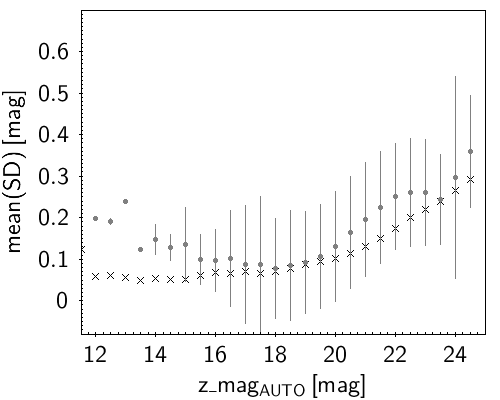}
\caption{Photometric repeatability as a function of magnitudes for sources observed more than five times for PSF (top) and AUTO (bottom) photometry.
Bullets are the average values of the standard deviations of magnitudes of repeated sources in bin sizes of 0.5\,mag and the error bars show their standard deviations.
Mean values of the average photometric errors shown in Figure~\ref{fig:sensitivity} are displayed as black crosses.
\label{fig:stdmag_mag}}
\end{figure*}

In order to estimate the internal photometric precision of the catalogue, we compared for each filter the magnitudes of sources observed more than five times.
The relation between their standard deviations and magnitudes is shown in Figure~\ref{fig:stdmag_mag}.
We also show in this figure the averaged photometric errors of Figure~\ref{fig:sensitivity}.
The scatter in the standard deviation of magnitudes increases towards fainter magnitudes in both, PSF and AUTO calibrated magnitudes. Also AUTO photometry shows slightly higher dispersion and error bars than PSF photometry. 
Since {\sc SExtractor} uses flexible elliptical apertures for extracting AUTO photometry, this effect can be attributed to the possibly different configurations applied to the same source in different images and epochs that lead to different flux integrations.
In general, the variations of magnitudes of repeated sources is of the order of or lower than the mean accuracy of the catalogue within the errorbars at each interval of magnitudes with the exception of the bright end of both, PSF and AUTO photometry, in the $g$ and $z$ bands. Also the faintest interval of magnitudes in the PSF panels in all bands and the AUTO photometry in the $i$ band show this behaviour, explained by the lower number of sources in this regime.

\begin{figure}
\centering
\includegraphics[width=0.45\textwidth]{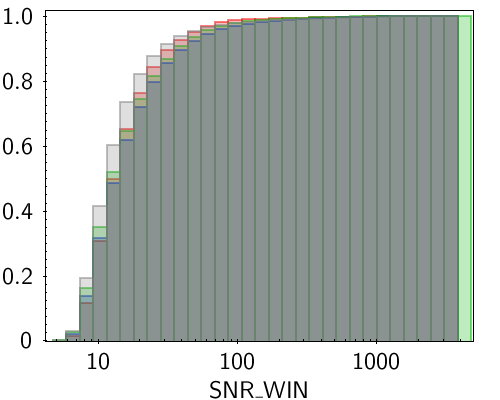}
\caption{Normalized cumulative distributions of the signal-to-noise ratios of sources observed more than five times and with large standard deviation of PSF magnitudes. The $g$, $r$, $i$ and $z$ filters are represented in red, blue, green and gray, respectively.
\label{fig:hist_rep}}
\end{figure}

While this exercise is a good approach to assess the homogeneity and quality of the entire dataset, it can also be misleading since the tails of the distributions are generally caused by faint sources with low signal-to-noise ratio and a small number of artifacts that passed previous tests. Figure~\ref{fig:hist_rep} shows, for each filter, the normalized cumulative histogram of the signal-to-noise ratio of sources with PSF magnitude standard deviations larger than the mean value plus 1$\sigma$ (this is, typically larger than 0.2\,mag). 
Per filter, between 75 and 85\% of these problematic sources have \verb|SNR_WIN| under 25.

\subsubsection{Variability}

\begin{figure}
\centering
\includegraphics[width=0.45\textwidth]{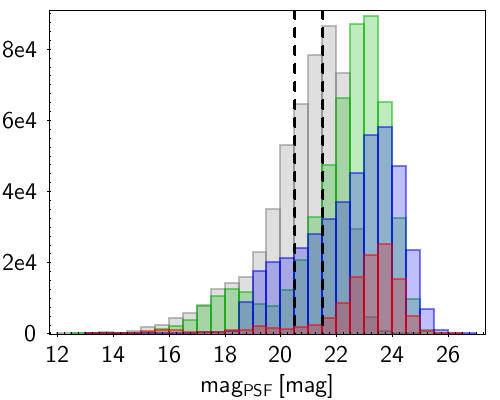}
\caption{PSF magnitude distributions of sources with $S$<0 in the $g$ (red), $r$ (blue), $i$ (green) and $z$ (gray) bands. Vertical black dashed lines indicate the 20.5--21.5\,mag limits of precise photometry.
\label{fig:hist_var}}
\end{figure}

Real variable objects can also contribute to the tail of the distribution.
We therefore looked for PSF magnitude variations according to the following equation:
\begin{equation}
   S= mag_{min}+3*e\_mag_{min}-(mag_{max}-3*e\_mag_{max})
\end{equation}

where $mag_{min}$, $mag_{max}$, $e\_mag_{min}$ and $e\_mag_{max}$ are the lower and higher magnitudes measured for the same source and their associated errors. If $S$ is lower than zero, then the magnitude deviation of the source is outside the reach of three times the errorbars and cannot be explained by the photometric errors.
Of the 731\,685 sources that have been detected more than once in the same filter we found that, between 10 and 13\%, fulfill this criterion, depending on the filter. Figure~\ref{fig:hist_var} shows the PSF magnitude distributions of these sources in the four bands. They peak well over the 20.5--21.5\,mag limits stated for precise photometry and therefore, these magnitude variations could be related in some cases to the lower quality photometry of faint sources rather than to real photometric variation. Nonetheless, 33\,310 out of the 731\,685 repeated sources (4.5\%) are brighter than those magnitude limits (21.5\,mag in $g$ and $r$, 21.0\,mag in $i$ and 20.5\,mag in $z$) and could actually be real photometric variable sources. Of them, only 16 have already been identified as variable in the literature. A proper analysis of the true nature of the variability detected in these sources is beyond the scope of this paper.

\begin{figure*}
\centering
\includegraphics[width=0.32\textwidth]{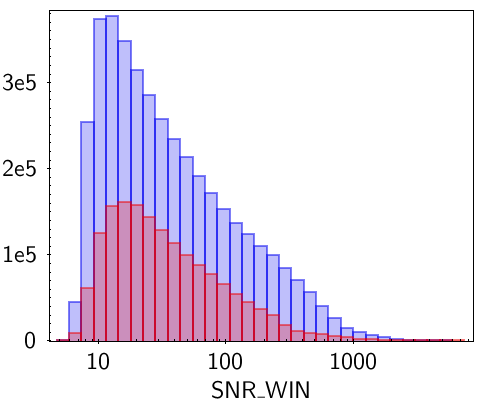}
\includegraphics[width=0.32\textwidth]{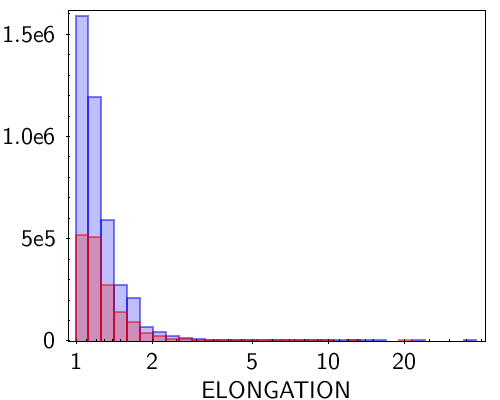}
\includegraphics[width=0.32\textwidth]{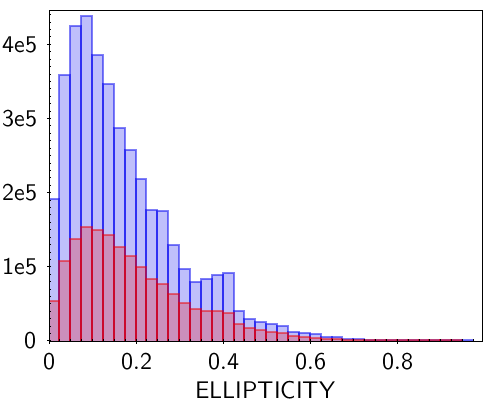}
\caption{Distribution of the SNR (left), elongation (middle) and ellipticity (right) of repeated sources in the catalogue with $S<0$ (red) and with $S>0$ (blue).
\label{fig:snr_dist_var}}
\end{figure*}

We also represent in Figure~\ref{fig:snr_dist_var} the distributions of the \verb|SNR_WIN|, elongation and ellipticity of the sources that show $S$ values smaller and greater than zero for comparison and in order to evaluate any drift of these parameters favouring variability.
Although sources with $S>0$ register a mean signal-to-noise ratio slightly higher compared to the mean value of sources with $S<0$ (85 vs. 64), the peak in both signal-to-noise ratio distributions is similar. The behaviour of morphometric parameters is comparable in both samples as well. Hence, we do not see any dependence of the photometric variability with lower signal-to-noise ratio or flattening of the sources.

\begin{figure}
\centering
\includegraphics[width=0.45\textwidth]{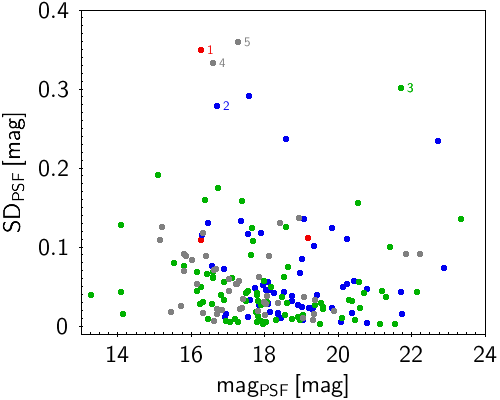}
\caption{Standard deviation of PSF magnitudes of known variable sources. Red, blue, green and gray filled circles stand for $g$, $r$, $i$ and $z$ bands respectively.
\label{fig:known_var}}
\end{figure}

On the other hand, we also investigated if known variable sources do also show variability in our catalogue.
To do so, we cross-matched our catalogue with {\it Gaia} DR2 and the AAVSO International Variable Star Index VSX \citep{2006SASS...25...47W}. Only 73 different sources in our catalogue are tagged as variables in these catalogues. We looked in the SIMBAD astronomical database \citep{Wenger00} and identified another 319 sources classified as variables or suspected of variability.
We show in Figure~\ref{fig:known_var} the PSF standard deviation of magnitudes for each of these sources with respect to their PSF magnitudes in the corresponding filter. The majority of these sources do not present significant variations except perhaps the five sources labeled in the plot with standard deviations over 0.2\,mag and larger than their magnitude errors.
Sources \#1 and \#2 correspond to the same source (NSV 13246) observed in the $g$ and $r$ bands, and classified in {\it Gaia} DR2 and SIMBAD as Fundamental-mode RR Lyrae star; source \#3 (SN 2011by) is a SN Ia in the AAVSO International Variable Star Index VSX catalogue; source \#4 (CRTS J154326.0-212800) is classified in {\it Gaia} DR2 and SIMBAD as a fundamental-mode RR Lyrae star; and source \#5 (2MASS J20325377+4115134) is found in SIMBAD as an eclipsing binary candidate, variable and X ray emitting source. Sources \#1 and \#2 have been detected within 0.83 years four and two times in our catalogue, respectively. Sources \#3, \#4 and \#5 have been detected twice in 56 days, six times within 10 days and six times within 18 days, respectively. Due to the time span of the observations, these magnitude variations could probably be associated to real photometric variations in all cases.
Unfortunately, the cadence in the observations might not be suitable for variability detection in the catalogue.

\section{Scientific exploitation}\label{sec:science}

To illustrate the science capabilities of the catalogue, we defined two cases to prove the goodness of the photometry (identification of cool stars) and the astrometry (identification of asteroids) provided in the catalogue.

\subsection{Looking for cool dwarfs}

Cool dwarfs of M spectral type constitute around the 40\% of the stellar mass in the Galaxy \citep{Gould96, Boc10}. 
Their ubiquity and lifetimes that exceed the current age of the Universe \citep{Bar98, Hen06} makes them excellent targets to broadly study the formation and evolution processes at the bottom of the main-sequence.

To identify cool dwarfs in the \osiris catalogue, we used the $i$ and $z$ bands and selected all sources with $i-z> 0.38$\,mag, which would correspond to spectral types later than M0 according to \cite{West08}. We then cross-matched the selected sources with {\it Gaia} DR2 within 1.5\,arcsec and kept all sources with relative errors in parallax and proper motion below 20\%. Later we removed all sources with proper motions $\mu < 30$\,mas/yr as in \cite{Solano19} to avoid contamination of giants and subgiants in the sample. This way, we ended up with 52 cool dwarf candidates, for which we derived effective temperatures in order to confirm their cool nature. To do so, we used VOSA \citep{Bayo08} which allowed us to gather photometry from the DENIS \citep{DENIS1999,DENIS2000}, 2MASS \citep{2MASS2006}, UKIDSS \citep{UKIDSS2006,UKIDSS2007,UKIDSS2007_2,UKIDSS2009}, IPHAS \citep{IPHAS2008,IPHAS2014}, WISE \citep{WISE2010}, SDSS DR9 \citep{SDSS2012}, Pan-STARRS DR1 \citep{Chambers16, Flewelling2016,Magnier2016, Magnier2016_1, Magnier2016_2, Tony12, Waters2016} and Gaia DR2 catalogues. This information was used together the photometric points in the $griz$ bands from our \osiris catalogue making use of the SVO Filter Profile Service to build the corresponding Spectral Energy Distributions (SEDs). We applied the BT-Settl collection of theoretical models \citep{BTSettl} with solar metallicity, $\log{g}$ between 4.5 and 6.0 and $T_{eff}$ between 1\,000 and 5\,000\,K. We left extinction $A_V$ as a free parameter varying from 0 to 1\,mag (assuming that extinction could be up to 1\,mag at 1\,kpc) since it can strongly modify the shape of the SED and, therefore, the parameters determination. VOSA fittings were visually inspected to confirm that \osiris photometry does not deviate from the SED.

The lack of good photometry prevented us from performing a reliable fit for three out of the 52 candidates.
One of these three turned out to be a non-catalogued resolved physical binary with a late-type primary separated 1.24\,arcsec at 195\,pc (241.7\,AU). Near a hundred binaries with late M/L dwarf primaries have been identified up to date \citep{Bouy03, Close03, Dupuy17, Gagliuffi15}. 2MASS, {\it Gaia} DR2 and \gtc \osiris are able to resolve the pair and we can therefore confirm common proper motion. We could estimate the spectral type of the components from the {\it Gaia} DR2 $G-RP$ colours and the updated version of Table~5 in \cite{PecautMamajek13}\footnote{\url{http://www.pas.rochester.edu/~emamajek/EEM_dwarf_UBVIJHK_colors_Teff.txt}}. The primary (GTC$\_$OSIRIS$\_$BBI$\_$DR1$\_$J203146.23+411437.0) would be an M9.5\,V with $G-RP$=1.627\,mag.
For the secondary (GTC$\_$OSIRIS$\_$BBI$\_$DR1$\_$J203146.19+411437.6), there is no RP photometry. From the $\Delta G = 1.0$\,mag of the system, we estimate the secondary to be an L1-L2 dwarf.
Another one of these three sources (GTC$\_$OSIRIS$\_$BBI$\_$DR1$\_$J181631.68+691152.9) has a close bright companion in our \osiris images at 2.3\,arcsec that saturates (and is therefore not in the catalogue) and that is not resolved by {\it Gaia} DR2, PS1, 2MASS or WISE. The combined photometry of the two sources in these catalogues prevent us from obtaining the SED of our cool dwarf candidate. We can not suggest nor discard physical binding due to the lack of available information.

\begin{figure}
\centering
\includegraphics[width=0.45\textwidth]{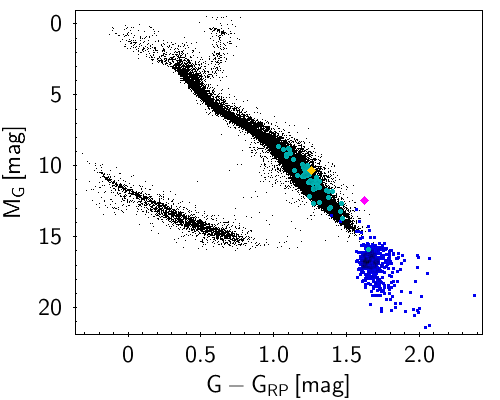}
\cprotect\caption{Colour-magnitude diagram using {\it Gaia} DR2 sources with parallaxes larger than 10\,mas (black dots). Light blue filled circles represent our 48 single dwarf candidates later than M0, yellow filled diamond represents the M type secondary belonging to the K+M close binary and the magenta filled diamond stands for the M+L close binary system. L and T dwarfs with {\it Gaia} counterparts identified in \cite{Smart17} are displayed with dark blue dots.}
\label{fig:hrd}
\end{figure}

In addition, we obtained effective temperatures for 49 sources ranging from 2\,400 to 3\,700 K with an uncertainty of 50\,K. Their positions shown in the colour-magnitude diagram of Figure~\ref{fig:hrd} built with {\it Gaia} DR2 sources also agree with being sources later than M0. Their distances range between 75.9 and 854.6\,pc. The coolest dwarf in the sample lies in the M/L transition region at 85.9\,pc. With an effective temperature of 2\,400\,K we estimate it's spectral type to be an M9.0-M9.5 dwarf. Again, using the {\it Gaia} $G-RP$ colour and the updated version of Table~5 in \cite{PecautMamajek13}, we infer a spectral type between M9.5 and L0.

Among the 49 cool sources, we found another non previously reported close binary separated 3.84\,arcsec (1\,940.4\,AU). We confirm common proper motion from the detection of both components by UKIDSS, Pan-STARRS DR1 and {\it Gaia} DR2. The pair is located at 505.3\,pc. In this work, we identified the secondary source of the system (GTC$\_$OSIRIS$\_$BBI$\_$DR1$\_$J203324.41+410751.7) and obtained an effective temperature of 3\,300\,K. For the primary (not included in the \osiris catalogue because of saturation in all images), we determined an effective temperature of 4\,200\,K using VOSA. From the {\it Gaia} DR2 $G-RP$ colours and the updated version of Table~5 in \cite{PecautMamajek13}, we estimated the spectral type of the components to be a K7 and an M4.0-M4.5. Spectral types are in agreement with the derived temperatures.

Only three out of the 49 sources were found in SIMBAD and none of them have been reported as cool type stars. As a curiosity, 36 out of the 49 sources cover less than one square degree of the sky in the region of the Cygnus OB association, although they are found at less than half its distance. This is a region of high scientific interest and may have therefore been often observed in several bands. This would explain the large fraction of cool objects found in this region.

Table~\ref{tab.cooldwarfs} lists the \osiris and {\it Gaia} DR2 identifiers and effective temperatures of the 49 dwarf candidates and the identifiers of the late-type binary components.

\subsection{Identification of asteroids}

We applied the \texttt{ssos}\footnote{\url{https://pypi.org/project/ssos/}} pipeline \citep{Max19} to detect and identify Solar System Objects (SSOs) serendipitously observed in the \osiris images. The pipeline detects both known and unknown SSOs primarily based on their linear apparent motion in subsequent exposures. Source detection and association are performed by {\sc SExtractor} and {\sc SCAMP} respectively, while the separation of SSOs from other sources in the image catalogues is performed by a chain of user-configurable filter algorithms. A more detailed discussion of the pipeline and its application to the \osiris images can be found in \cite{Max19}. 

To apply the pipeline, the images were grouped by observation night and overlapping field-of-views. Of the full sample of images in the DR1, we built 420 groups made up of 6\,982 images. The remaining images had to be discarded as there were fewer than 4 exposures in the respective visits, a requirement for a reliable detection of SSOs.

\begin{figure}
    \centering
    \includegraphics[width=0.45\textwidth]{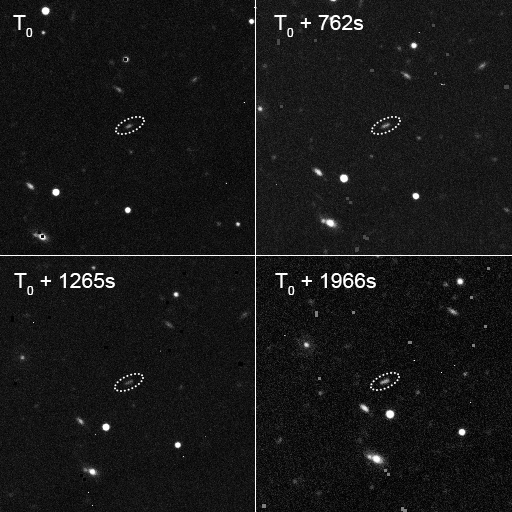}
    \caption{Four observations of \textit{ (355891) 2008 WE46} recovered from the DR1 images. The main-belt asteroid was serendipitously observed and recovered using the \textit{ssos} pipeline. Its position is marked by the dotted white ellipse in the center of each frame. The text indicates the time difference between the frames in seconds.}
    \label{fig:sso_example}
\end{figure}

Applying the \texttt{ssos} pipeline to these 420 groups revealed 204 unique SSOs present in a total of 2\,828 images.  63 objects could be identified as known SSOs using the IMCCE's SkyBoT service, which computes the ephemerides of SSOs within a given  field-of-view and observation epoch. The returned computed ephemerides were cross-matched within a radius of 40\,arcsec with the positions of the recovered SSOs. Table~\ref{table:skybot_gtc} lists the classes of the 63 identified objects. The majority are Main-Belt (MB) asteroids. Four comets were retrieved as well, however, they were the targets of the respective observations. Figure~\ref{fig:sso_example} depicts four detections of \textit{ (355891) 2008 WE46} to illustrate the serendipitous observations.

The remaining 141 SSOs are either unknown or had a discrepancy between predicted and observed position larger than 40\,arcsec, meaning that their observation will greatly improve the accuracy of their orbit. The astrometric and photometric properties of all 2\,828 SSO detections have been reported to the Minor Planet Centre\footnote{\url{https://www.minorplanetcenter.net/}} (MPC). Of the 1\,002 observations of known SSOs, 872 were ingested into the MPC database\footnote{\url{https://minorplanetcenter.net/iau/ECS/MPCArchive/2018/MPS_20181118.pdf}, observatory code Z18}. The remaining 130 observations have not been published due to unknown reasons.
The 1\,826 observations of unknown SSOs consist of single-night observations only. Therefore, they will not receive temporary designations, however, they are ingested into the MPC database and might eventually be associated to a newly discovered object.

Unfortunately, the acquired photometry did not allow for spectrophotometric classification of the SSOs as the observations typically spanned more than 1 hour, rendering the determination of colours unreliable. Moreover, the temporal baseline was not long enough to estimate rotation periods from the light curve analysis. The observations will still be useful in combination with other data, however, to determine e.g. the SSO phase function parameters.

\begin{table}
        \centering
        \caption {Distribution of previously known SSOs detected in the DR1 images over SSO classes. The 4 comets were retrieved from targeted observations. MB stands for Main-Belt.}
        \label{table:skybot_gtc}
        \begin{tabular}{lllll}
        \hline \hline
        \noalign{\smallskip}
Comet  & Inner MB &Middle MB &Outer MB & Trojan\\
\noalign{\smallskip}
        \hline
4  & 21 & 17 & 20 & 1 \\
        \noalign{\smallskip}
        \hline
        \end{tabular}
\end{table}

\section{Data Access}\label{sec:data_access}

The photometrically and astrometrically corrected \osiris broadband images as well as the associated catalogue are available to the community through the \gtc Archive Portal\footnote{\url{http://gtc.sdc.cab.inta-csic.es}} or the associated Virtual Observatory services (SIAP for images and ConeSearch for the catalogue). The \gtc archive is maintained by Centro de Astrobiologia (INTA-CSIC) in the framework of the Spanish Virtual Observatory\footnote{\url{http://svo.cab.inta-csic.es}}. The results provided by the portal or the VO services can be sent using the SAMP protocol to other VO tools for its further visualization and/or analysis. 

In order to help the astronomical community on using the detection and source catalogues built from the \osiris broadband images, we have developed an archive system that can be accessed from a webpage\footnote{\url{http://svo2.cab.inta-csic.es/vocats/v2/gtc-osiris/}} or through a Virtual Observatory ConeSearch\footnote{e.g. \url{http://svo2.cab.inta-csic.es/vocats/v2/gtc-osiris-primary/cs.php?RA=0.107&DEC=44.636&SR=0.1&VERB=2}}.

\begin{figure*}
        \centering
        \includegraphics[width=\hsize]{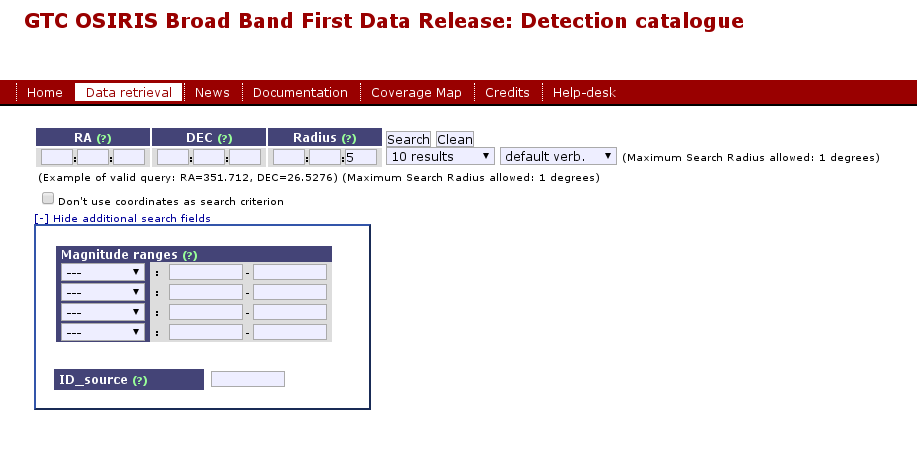}
        \includegraphics[width=\hsize]{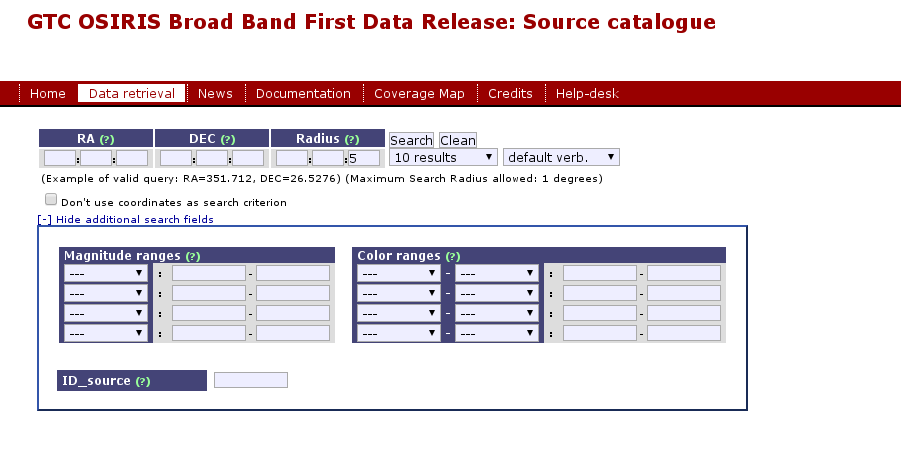}
        \caption{Screenshots of the archive search interfaces that permit simple queries to the detection (top) and source (bottom) catalogues.}
         \label{fig.svoquery}
  \end{figure*}

The archive system implements a very simple search interface (see Fig~\ref{fig.svoquery}) that permits queries by position or PSF magnitude interval in both, detection and source catalogues, and also by colour range only in the source catalogue. The system implements aswell a link to the images in \gtc Public Archive.

The result of the query is a HTML table with all the sources found in the archive fulfilling the search criteria up to a limit of 10\,000 lines. The result can also be downloaded as a VOTable or a CSV file. Detailed information on the output fields can be obtained placing the mouse over the question mark (``?") located close to the name of the column. The archive also implements the SAMP\footnote{\url{http://www.ivoa.net/documents/SAMP/}} (Simple Application Messaging) Virtual Observatory protocol. SAMP allows Virtual Observatory applications to communicate with each other in a seamless and transparent manner for the user. This way, the results of a query can be easily transferred to other VO applications, such as, for instance, TOPCAT.

\section{Conclusions and future work}\label{sec:conclusions}

We have presented the database of processed and scientific grade broadband images and the associated catalogue obtained with the \gtc \osiris instrument from April 2009 to January 2014. In this first release, the database includes 6\,788 images in the Sloan $griz$ bands with exposure times ranging from 0.5 to 900\,s, and 6\,226\,520 entries in the catalogue corresponding to 633\,559 astronomical sources. The catalogue is astrometrically and photometrically calibrated in standard PSF and automatic (Kron equivalent) photometry using 2MASS, SDSS DR10 or USNO-B1 for the astrometry and Pan-STARRS DR1 for the photometry, allowing a broad range of scientific activities. 

Relative astrometric residuals typically are within 30\,mas and typical positional uncertainty is of 0.12\,arcsec. Only well correlated sources between instrumental and Pan-STARRS DR1 magnitudes were used for photometric calibration, providing a precision of 0.034--0.056\,mag, depending on the filter.

The mean photometric accuracy of the whole catalogue is better than 0.09\,mag and 0.15\,mag in PSF and AUTO photometry, respectively. Besides, a comparison with PAN-STARRS DR1 good quality photometry presents magnitude differences under 0.18\,mag in all bands.
Additional tests did not reveal any colour term in the photometric calibration nor dependence with the binning mode of observations nor the pixel position in the CCD.

Saturation and typical magnitude limits in the catalogue in each filter are 13.1--24.6\,mag in $g$, 13.6--24.5\,mag in $r$, 13.1--23.9\,mag in $i$, and 12.0--22.7\,mag in $z$, respectively.

We present two science cases aiming to prove the scientific capabilities of the catalogue. In the first case, we looked for cool dwarfs in the catalogue using a photometric criterium in the optical and {\it Gaia} DR2 astrometry. We identified 49 dwarfs of spectral types later than M0 according to their positions in the colour-magnitude diagram and their effective temperatures, with distances between 75.9 and 854.6\,pc, which need spectroscopic confirmation. Among them, we identified a new resolved binary system separated 3.84\,arcsec (1\,940.4\,AU) composed by a late K primary and a mid type M secondary. Additionally, we identified another new resolved binary system separated 1.24\,arcsec (247.7\,AU) with estimated spectral types M9.5 + L1/L2 dwarfs. We propose them for follow up and spectroscopic confirmation.
In the second case, we looked for Solar System Objects from their apparent linear motion and identified 59 known asteroids and four known comets plus 141 unknown objects or objects with poor accuracy in their orbital elements. For them, we provide 2\,828 detections that have been reported to the Minor Planet Center Database.

The catalogue includes celestial coordinates, sources and detections IDs, calibrated PSF and Kron photometry in the $griz$ Sloan bands, various morphometric measurements useful to assess the nature of the sources (point-like or extended), although we provide a point-like/extended classification and a point-like coefficient.
In addition, it contains two quality flags associated to the photometric coverage of the calibration using PAN-STARRS DR1 catalogue.
Instrumental magnitudes of \verb|MODEL| and aperture photometry, and epoch of observation are also included in the catalogue.
The user can access the raw and processed images through an url given in the catalogue.

New releases of the catalogue including the new public broadband images will be delivered over the life of the instrument. Also, a number of improvements and enhancements are planned for these future releases. One of the most important ones will be the building of stacked images to develop a much deeper catalogue. We also plan to improve the absolute astrometry of the catalogue by linking it to the {\it Gaia} DR2 catalogue. New versions of the Pan-Starrs will also be used for the photometric calibration.

Like in any other photometric survey, the complete absence of errors and problems in the catalogue cannot be guaranteed, in particular for detections close to the limiting magnitude. In these cases, users are strongly encouraged to download and check the associated images to assess the reliability of a given catalogue measurement. The {\it News} section of the catalogue website will contain a list of Frequently Asked Questions as well as a description of caveats that may arise with the scientific exploitation of the catalogue.

\section*{Acknowledgments}
H. Bouy acknowledges funding from the European Research Council (ERC) under the European Union's Horizon 2020 research and innovation programme (grant agreement No 682903, P.I. H. Bouy), and from the French State in the framework of the "Investments for the future" Program, IdEx Bordeaux, reference ANR-10-IDEX-03-02.
This research has been financed by ASTERICS, a project supported by the European Commission Framework Programme Horizon 2020 Research and Innovation action under grant agreement n. 653477.
This research has been partially funded by the Spanish State Research Agency (AEI) Project No. ESP2017-87676-C5-1-R and No. MDM-2017-0737 Unidad de Excelencia ``Mar\'ia de Maeztu''- Centro de Astrobiolog\'ia (CSIC-INTA).
This publication makes use of VOSA, developed under the Spanish Virtual Observatory project supported by the Spanish MINECO through grant AyA2017-84089.
This research has made use of the SVO Filter Profile Service (http://svo2.cab.inta-csic.es/theory/fps/) supported from the Spanish MINECO through grant AYA2017-84089.
This research made use of the cross-match service, the SIMBAD database \citep{Wenger00}, VizieR catalogue access tool \citep{Och00}, "Aladin sky atlas" \citep{Bonn00, BF14} provided by CDS, Strasbourg, France.
This research has also made use of the TOPCAT \citep{Taylor05} and STILTS \citep{Taylor06}.
MCC and FJE acknowledge financial support from the Tec2Space-CM project (P2018/NMT-4291).
MM is funded by the European Space Agency under the research contract C4000122918.

\bibliographystyle{mnras}
\bibliography{biblio.bib}

\begin{thebibliography}{}
\makeatletter
\relax
\def\mn@urlcharsother{\let\do\@makeother \do\$\do\&\do\#\do\^\do\_\do\%\do\~}
\def\mn@doi{\begingroup\mn@urlcharsother \@ifnextchar [ {\mn@doi@}
  {\mn@doi@[]}}
\def\mn@doi@[#1]#2{\def\@tempa{#1}\ifx\@tempa\@empty \href
  {http://dx.doi.org/#2} {doi:#2}\else \href {http://dx.doi.org/#2} {#1}\fi
  \endgroup}
\def\mn@eprint#1#2{\mn@eprint@#1:#2::\@nil}
\def\mn@eprint@arXiv#1{\href {http://arxiv.org/abs/#1} {{\tt arXiv:#1}}}
\def\mn@eprint@dblp#1{\href {http://dblp.uni-trier.de/rec/bibtex/#1.xml}
  {dblp:#1}}
\def\mn@eprint@#1:#2:#3:#4\@nil{\def\@tempa {#1}\def\@tempb {#2}\def\@tempc
  {#3}\ifx \@tempc \@empty \let \@tempc \@tempb \let \@tempb \@tempa \fi \ifx
  \@tempb \@empty \def\@tempb {arXiv}\fi \@ifundefined
  {mn@eprint@\@tempb}{\@tempb:\@tempc}{\expandafter \expandafter \csname
  mn@eprint@\@tempb\endcsname \expandafter{\@tempc}}}

\bibitem[\protect\citeauthoryear{{Ahn} et~al.,}{{Ahn} et~al.}{2012}]{SDSS2012}
{Ahn} C.~P.,  et~al., 2012, \mn@doi [\apjs] {10.1088/0067-0049/203/2/21}, \href
  {http://cdsads.u-strasbg.fr/abs/2012ApJS..203...21A} {203, 21}

\bibitem[\protect\citeauthoryear{{Ahn} et~al.,}{{Ahn}
  et~al.}{2014}]{2014ApJS..211...17A}
{Ahn} C.~P.,  et~al., 2014, \mn@doi [\apjs] {10.1088/0067-0049/211/2/17}, \href
  {http://adsabs.harvard.edu/abs/2014ApJS..211...17A} {211, 17}

\bibitem[\protect\citeauthoryear{{Allard}, {Homeier}  \& {Freytag}}{{Allard}
  et~al.}{2012}]{BTSettl}
{Allard} F.,  {Homeier} D.,   {Freytag} B.,  2012, \mn@doi [Philosophical
  Transactions of the Royal Society of London Series A]
  {10.1098/rsta.2011.0269}, \href
  {http://adsabs.harvard.edu/abs/2012RSPTA.370.2765A} {370, 2765}

\bibitem[\protect\citeauthoryear{{Baraffe}, {Chabrier}, {Allard}  \&
  {Hauschildt}}{{Baraffe} et~al.}{1998}]{Bar98}
{Baraffe} I.,  {Chabrier} G.,  {Allard} F.,   {Hauschildt} P.~H.,  1998, \aap,
  \href {http://adsabs.harvard.edu/abs/1998A%26A...337..403B} {337, 403}

\bibitem[\protect\citeauthoryear{{Bardalez Gagliuffi}, {Gelino}  \&
  {Burgasser}}{{Bardalez Gagliuffi} et~al.}{2015}]{Gagliuffi15}
{Bardalez Gagliuffi} D.~C.,  {Gelino} C.~R.,   {Burgasser} A.~J.,  2015,
  \mn@doi [\aj] {10.1088/0004-6256/150/5/163}, \href
  {https://ui.adsabs.harvard.edu/abs/2015AJ....150..163B} {150, 163}

\bibitem[\protect\citeauthoryear{{Barentsen} et~al.,}{{Barentsen}
  et~al.}{2014}]{IPHAS2014}
{Barentsen} G.,  et~al., 2014, \mn@doi [\mnras] {10.1093/mnras/stu1651}, \href
  {http://adsabs.harvard.edu/abs/2014MNRAS.444.3230B} {444, 3230}

\bibitem[\protect\citeauthoryear{{Bayo}, {Rodrigo}, {Barrado Y Navascu{\'e}s},
  {Solano}, {Guti{\'e}rrez}, {Morales-Calder{\'o}n}  \& {Allard}}{{Bayo}
  et~al.}{2008}]{Bayo08}
{Bayo} A.,  {Rodrigo} C.,  {Barrado Y Navascu{\'e}s} D.,  {Solano} E.,
  {Guti{\'e}rrez} R.,  {Morales-Calder{\'o}n} M.,   {Allard} F.,  2008, \mn@doi
  [\aap] {10.1051/0004-6361:200810395}, \href
  {http://adsabs.harvard.edu/abs/2008A%26A...492..277B} {492, 277}

\bibitem[\protect\citeauthoryear{{Bertin}}{{Bertin}}{2006}]{2006ASPC..351..112B}
{Bertin} E.,  2006, in {Gabriel} C.,  {Arviset} C.,  {Ponz} D.,   {Enrique} S.,
   eds,  Astronomical Society of the Pacific Conference Series Vol. 351,
  Astronomical Data Analysis Software and Systems XV. p.~112

\bibitem[\protect\citeauthoryear{{Bertin}}{{Bertin}}{2013}]{2013ascl.soft01001B}
{Bertin} E.,  2013, {PSFEx: Point Spread Function Extractor}, Astrophysics
  Source Code Library (\mn@eprint {ascl} {1301.001})

\bibitem[\protect\citeauthoryear{{Bertin} \& {Arnouts}}{{Bertin} \&
  {Arnouts}}{1996}]{1996A&AS..117..393B}
{Bertin} E.,  {Arnouts} S.,  1996, Astronomy \& Astrophysics Supplement, \href
  {http://adsabs.harvard.edu/abs/1996A%26AS..117..393B} {117, 393}

\bibitem[\protect\citeauthoryear{{Bertin}, {Mellier}, {Radovich}, {Missonnier},
  {Didelon}  \& {Morin}}{{Bertin} et~al.}{2002}]{2002ASPC..281..228B}
{Bertin} E.,  {Mellier} Y.,  {Radovich} M.,  {Missonnier} G.,  {Didelon} P.,
  {Morin} B.,  2002, in {Bohlender} D.~A.,  {Durand} D.,   {Handley} T.~H.,
  eds,  Astronomical Society of the Pacific Conference Series Vol. 281,
  Astronomical Data Analysis Software and Systems XI. p.~228

\bibitem[\protect\citeauthoryear{{Boch} \& {Fernique}}{{Boch} \&
  {Fernique}}{2014}]{BF14}
{Boch} T.,  {Fernique} P.,  2014, in {Manset} N.,  {Forshay} P.,  eds,
  Astronomical Society of the Pacific Conference Series Vol. 485, Astronomical
  Data Analysis Software and Systems XXIII. p.~277

\bibitem[\protect\citeauthoryear{{Bochanski}, {Hawley}, {Covey}, {West},
  {Reid}, {Golimowski}  \& {Ivezi{\'c}}}{{Bochanski} et~al.}{2010}]{Boc10}
{Bochanski} J.~J.,  {Hawley} S.~L.,  {Covey} K.~R.,  {West} A.~A.,  {Reid}
  I.~N.,  {Golimowski} D.~A.,   {Ivezi{\'c}} {\v Z}.,  2010, \mn@doi [\aj]
  {10.1088/0004-6256/139/6/2679}, \href
  {http://adsabs.harvard.edu/abs/2010AJ....139.2679B} {139, 2679}

\bibitem[\protect\citeauthoryear{{Bonnarel} et~al.,}{{Bonnarel}
  et~al.}{2000}]{Bonn00}
{Bonnarel} F.,  et~al., 2000, \mn@doi [\aaps] {10.1051/aas:2000331}, \href
  {http://adsabs.harvard.edu/abs/2000A%26AS..143...33B} {143, 33}

\bibitem[\protect\citeauthoryear{{Bouy}, {Brandner}, {Mart{\'\i}n}, {Delfosse},
  {Allard}  \& {Basri}}{{Bouy} et~al.}{2003}]{Bouy03}
{Bouy} H.,  {Brandner} W.,  {Mart{\'\i}n} E.~L.,  {Delfosse} X.,  {Allard} F.,
   {Basri} G.,  2003, \mn@doi [\aj] {10.1086/377343}, \href
  {https://ui.adsabs.harvard.edu/abs/2003AJ....126.1526B} {126, 1526}

\bibitem[\protect\citeauthoryear{{Bouy}, {Bertin}, {Moraux}, {Cuillandre},
  {Bouvier}, {Barrado}, {Solano}  \& {Bayo}}{{Bouy}
  et~al.}{2013}]{2013A&A...554A.101B}
{Bouy} H.,  {Bertin} E.,  {Moraux} E.,  {Cuillandre} J.-C.,  {Bouvier} J.,
  {Barrado} D.,  {Solano} E.,   {Bayo} A.,  2013, \mn@doi [Astronomy \&
  Astrophysics] {10.1051/0004-6361/201220748}, \href
  {http://cdsads.u-strasbg.fr/abs/2013A%26A...554A.101B} {554, A101}

\bibitem[\protect\citeauthoryear{{Casali} et~al.,}{{Casali}
  et~al.}{2007}]{UKIDSS2007}
{Casali} M.,  et~al., 2007, \mn@doi [\aap] {10.1051/0004-6361:20066514}, \href
  {http://adsabs.harvard.edu/abs/2007A%26A...467..777C} {467, 777}

\bibitem[\protect\citeauthoryear{{Chambers} et~al.,}{{Chambers}
  et~al.}{2016}]{Chambers16}
{Chambers} K.~C.,  et~al., 2016, preprint, \href
  {http://cdsads.u-strasbg.fr/abs/2016arXiv161205560C} {} (\mn@eprint {arXiv}
  {1612.05560})

\bibitem[\protect\citeauthoryear{{Close}, {Siegler}, {Freed}  \&
  {Biller}}{{Close} et~al.}{2003}]{Close03}
{Close} L.~M.,  {Siegler} N.,  {Freed} M.,   {Biller} B.,  2003, \mn@doi [\apj]
  {10.1086/368177}, \href
  {https://ui.adsabs.harvard.edu/abs/2003ApJ...587..407C} {587, 407}

\bibitem[\protect\citeauthoryear{{Cutri} et~al.,}{{Cutri}
  et~al.}{2003}]{2003yCat.2246....0C}
{Cutri} R.~M.,  et~al., 2003, VizieR Online Data Catalog, \href
  {http://cdsads.u-strasbg.fr/abs/2003yCat.2246....0C} {2246, 0}

\bibitem[\protect\citeauthoryear{{Dupuy} \& {Liu}}{{Dupuy} \&
  {Liu}}{2017}]{Dupuy17}
{Dupuy} T.~J.,  {Liu} M.~C.,  2017, \mn@doi [\apjs] {10.3847/1538-4365/aa5e4c},
  \href {https://ui.adsabs.harvard.edu/abs/2017ApJS..231...15D} {231, 15}

\bibitem[\protect\citeauthoryear{{Epchtein} et~al.,}{{Epchtein}
  et~al.}{1999}]{DENIS1999}
{Epchtein} N.,  et~al., 1999, \aap, \href
  {http://adsabs.harvard.edu/abs/1999A%26A...349..236E} {349, 236}

\bibitem[\protect\citeauthoryear{{Farrow} et~al.,}{{Farrow}
  et~al.}{2014}]{Farrow2014}
{Farrow} D.~J.,  et~al., 2014, \mn@doi [\mnras] {10.1093/mnras/stt1933}, \href
  {http://adsabs.harvard.edu/abs/2014MNRAS.437..748F} {437, 748}

\bibitem[\protect\citeauthoryear{{Flewelling} et~al.,}{{Flewelling}
  et~al.}{2016}]{Flewelling2016}
{Flewelling} H.~A.,  et~al., 2016, arXiv e-prints, \href
  {https://ui.adsabs.harvard.edu/abs/2016arXiv161205243F} {p. arXiv:1612.05243}

\bibitem[\protect\citeauthoryear{{Fouqu{\'e}} et~al.,}{{Fouqu{\'e}}
  et~al.}{2000}]{DENIS2000}
{Fouqu{\'e}} P.,  et~al., 2000, \mn@doi [\aaps] {10.1051/aas:2000123}, \href
  {http://adsabs.harvard.edu/abs/2000A%26AS..141..313F} {141, 313}

\bibitem[\protect\citeauthoryear{{Gonz{\'a}lez-Solares}
  et~al.,}{{Gonz{\'a}lez-Solares} et~al.}{2008}]{IPHAS2008}
{Gonz{\'a}lez-Solares} E.~A.,  et~al., 2008, \mn@doi [\mnras]
  {10.1111/j.1365-2966.2008.13399.x}, \href
  {http://adsabs.harvard.edu/abs/2008MNRAS.388...89G} {388, 89}

\bibitem[\protect\citeauthoryear{{Gould}, {Bahcall}  \& {Flynn}}{{Gould}
  et~al.}{1996}]{Gould96}
{Gould} A.,  {Bahcall} J.~N.,   {Flynn} C.,  1996, \mn@doi [\apj]
  {10.1086/177460}, \href {http://adsabs.harvard.edu/abs/1996ApJ...465..759G}
  {465, 759}

\bibitem[\protect\citeauthoryear{{Henry}, {Jao}, {Subasavage}, {Beaulieu},
  {Ianna}, {Costa}  \& {M{\'e}ndez}}{{Henry} et~al.}{2006}]{Hen06}
{Henry} T.~J.,  {Jao} W.-C.,  {Subasavage} J.~P.,  {Beaulieu} T.~D.,  {Ianna}
  P.~A.,  {Costa} E.,   {M{\'e}ndez} R.~A.,  2006, \mn@doi [\aj]
  {10.1086/508233}, \href {http://adsabs.harvard.edu/abs/2006AJ....132.2360H}
  {132, 2360}

\bibitem[\protect\citeauthoryear{{Hewett}, {Warren}, {Leggett}  \&
  {Hodgkin}}{{Hewett} et~al.}{2006}]{UKIDSS2006}
{Hewett} P.~C.,  {Warren} S.~J.,  {Leggett} S.~K.,   {Hodgkin} S.~T.,  2006,
  \mn@doi [\mnras] {10.1111/j.1365-2966.2005.09969.x}, \href
  {http://adsabs.harvard.edu/abs/2006MNRAS.367..454H} {367, 454}

\bibitem[\protect\citeauthoryear{{Hodgkin}, {Irwin}, {Hewett}  \&
  {Warren}}{{Hodgkin} et~al.}{2009}]{UKIDSS2009}
{Hodgkin} S.~T.,  {Irwin} M.~J.,  {Hewett} P.~C.,   {Warren} S.~J.,  2009,
  \mn@doi [\mnras] {10.1111/j.1365-2966.2008.14387.x}, \href
  {http://adsabs.harvard.edu/abs/2009MNRAS.394..675H} {394, 675}

\bibitem[\protect\citeauthoryear{{Kaiser} et~al.,}{{Kaiser}
  et~al.}{2010}]{Kaiser10}
{Kaiser} N.,  et~al., 2010, The Pan-STARRS wide-field optical/NIR imaging
  survey, \mn@doi{10.1117/12.859188}, \url {https://doi.org/10.1117/12.859188}

\bibitem[\protect\citeauthoryear{{Lawrence} et~al.,}{{Lawrence}
  et~al.}{2007a}]{Lawrence07}
{Lawrence} A.,  et~al., 2007a, \mn@doi [\mnras]
  {10.1111/j.1365-2966.2007.12040.x}, \href
  {http://adsabs.harvard.edu/abs/2007MNRAS.379.1599L} {379, 1599}

\bibitem[\protect\citeauthoryear{{Lawrence} et~al.,}{{Lawrence}
  et~al.}{2007b}]{UKIDSS2007_2}
{Lawrence} A.,  et~al., 2007b, \mn@doi [\mnras]
  {10.1111/j.1365-2966.2007.12040.x}, \href
  {http://adsabs.harvard.edu/abs/2007MNRAS.379.1599L} {379, 1599}

\bibitem[\protect\citeauthoryear{{Magnier} et~al.,}{{Magnier}
  et~al.}{2016a}]{Magnier2016}
{Magnier} E.~A.,  et~al., 2016a, preprint, \href
  {http://adsabs.harvard.edu/abs/2016arXiv161205242M} {} (\mn@eprint {arXiv}
  {1612.05242})

\bibitem[\protect\citeauthoryear{{Magnier} et~al.,}{{Magnier}
  et~al.}{2016b}]{Magnier2016_1}
{Magnier} E.~A.,  et~al., 2016b, arXiv e-prints, \href
  {https://ui.adsabs.harvard.edu/abs/2016arXiv161205240M} {p. arXiv:1612.05240}

\bibitem[\protect\citeauthoryear{{Magnier} et~al.,}{{Magnier}
  et~al.}{2016c}]{Magnier2016_2}
{Magnier} E.~A.,  et~al., 2016c, arXiv e-prints, \href
  {https://ui.adsabs.harvard.edu/abs/2016arXiv161205244M} {p. arXiv:1612.05244}

\bibitem[\protect\citeauthoryear{{Mahlke}, {Solano}, {Bouy}, {Carry}, {Verdoes
  Kleijn}  \& {Bertin}}{{Mahlke} et~al.}{2019}]{Max19}
{Mahlke} M.,  {Solano} E.,  {Bouy} H.,  {Carry} B.,  {Verdoes Kleijn} G.~A.,
  {Bertin} E.,  2019, \mn@doi [Astronomy and Computing]
  {10.1016/j.ascom.2019.100289}, \href
  {https://ui.adsabs.harvard.edu/abs/2019A&C....2800289M} {28, 100289}

\bibitem[\protect\citeauthoryear{{Monet} et~al.,}{{Monet}
  et~al.}{2003}]{Monet03}
{Monet} D.~G.,  et~al., 2003, \mn@doi [\aj] {10.1086/345888}, \href
  {http://adsabs.harvard.edu/abs/2003AJ....125..984M} {125, 984}

\bibitem[\protect\citeauthoryear{{Ochsenbein}, {Bauer}  \&
  {Marcout}}{{Ochsenbein} et~al.}{2000}]{Och00}
{Ochsenbein} F.,  {Bauer} P.,   {Marcout} J.,  2000, \mn@doi [\aaps]
  {10.1051/aas:2000169}, \href
  {http://adsabs.harvard.edu/abs/2000A%26AS..143...23O} {143, 23}

\bibitem[\protect\citeauthoryear{{Paillassa}, {Bertin}  \& {Bouy}}{{Paillassa}
  et~al.}{2019}]{maximask}
{Paillassa} M.,  {Bertin} E.,   {Bouy} H.,  2019, arXiv e-prints, \href
  {https://ui.adsabs.harvard.edu/abs/2019arXiv190708298P} {p. arXiv:1907.08298}

\bibitem[\protect\citeauthoryear{{Pecaut} \& {Mamajek}}{{Pecaut} \&
  {Mamajek}}{2013}]{PecautMamajek13}
{Pecaut} M.~J.,  {Mamajek} E.~E.,  2013, \mn@doi [\apjs]
  {10.1088/0067-0049/208/1/9}, \href
  {https://ui.adsabs.harvard.edu/abs/2013ApJS..208....9P} {208, 9}

\bibitem[\protect\citeauthoryear{{Skrutskie} et~al.,}{{Skrutskie}
  et~al.}{2006}]{2MASS2006}
{Skrutskie} M.~F.,  et~al., 2006, \mn@doi [\aj] {10.1086/498708}, \href
  {http://adsabs.harvard.edu/abs/2006AJ....131.1163S} {131, 1163}

\bibitem[\protect\citeauthoryear{{Smart}, {Marocco}, {Caballero}, {Jones},
  {Barrado}, {Beam{\'\i}n}, {Pinfield}  \& {Sarro}}{{Smart}
  et~al.}{2017}]{Smart17}
{Smart} R.~L.,  {Marocco} F.,  {Caballero} J.~A.,  {Jones} H.~R.~A.,  {Barrado}
  D.,  {Beam{\'\i}n} J.~C.,  {Pinfield} D.~J.,   {Sarro} L.~M.,  2017, \mn@doi
  [\mnras] {10.1093/mnras/stx800}, \href
  {https://ui.adsabs.harvard.edu/abs/2017MNRAS.469..401S} {469, 401}

\bibitem[\protect\citeauthoryear{{Solano} et~al.,}{{Solano}
  et~al.}{2019}]{Solano19}
{Solano} E.,  et~al., 2019, \mn@doi [\aap] {10.1051/0004-6361/201935256}, \href
  {https://ui.adsabs.harvard.edu/abs/2019A&A...627A..29S} {627, A29}

\bibitem[\protect\citeauthoryear{{Taylor}}{{Taylor}}{2005}]{Taylor05}
{Taylor} M.~B.,  2005, in {Shopbell} P.,  {Britton} M.,   {Ebert} R.,  eds,
  Astronomical Society of the Pacific Conference Series Vol. 347, Astronomical
  Data Analysis Software and Systems XIV. p.~29

\bibitem[\protect\citeauthoryear{{Taylor}}{{Taylor}}{2006}]{Taylor06}
{Taylor} M.~B.,  2006, in {Gabriel} C.,  {Arviset} C.,  {Ponz} D.,   {Enrique}
  S.,  eds,  Astronomical Society of the Pacific Conference Series Vol. 351,
  Astronomical Data Analysis Software and Systems XV. p.~666

\bibitem[\protect\citeauthoryear{{Tonry} et~al.,}{{Tonry}
  et~al.}{2012}]{Tony12}
{Tonry} J.~L.,  et~al., 2012, \mn@doi [\apj] {10.1088/0004-637X/750/2/99},
  \href {https://ui.adsabs.harvard.edu/abs/2012ApJ...750...99T} {750, 99}

\bibitem[\protect\citeauthoryear{{Vandame}}{{Vandame}}{2002}]{2002SPIE.4847..123V}
{Vandame} B.,  2002, in {Starck} J.-L.,  {Murtagh} F.~D.,  eds,  Society of
  Photo-Optical Instrumentation Engineers (SPIE) Conference Series Vol. 4847,
  Astronomical Data Analysis II. pp 123--134 (\mn@eprint {}
  {astro-ph/0208230}), \mn@doi{10.1117/12.460591}

\bibitem[\protect\citeauthoryear{{Waters} et~al.,}{{Waters}
  et~al.}{2016}]{Waters2016}
{Waters} C.~Z.,  et~al., 2016, arXiv e-prints, \href
  {https://ui.adsabs.harvard.edu/abs/2016arXiv161205245W} {p. arXiv:1612.05245}

\bibitem[\protect\citeauthoryear{{Watson}, {Henden}  \& {Price}}{{Watson}
  et~al.}{2006}]{2006SASS...25...47W}
{Watson} C.~L.,  {Henden} A.~A.,   {Price} A.,  2006, Society for Astronomical
  Sciences Annual Symposium, \href
  {http://adsabs.harvard.edu/abs/2006SASS...25...47W} {25, 47}

\bibitem[\protect\citeauthoryear{{Wenger} et~al.,}{{Wenger}
  et~al.}{2000}]{Wenger00}
{Wenger} M.,  et~al., 2000, \mn@doi [\aaps] {10.1051/aas:2000332}, \href
  {http://adsabs.harvard.edu/abs/2000A%26AS..143....9W} {143, 9}

\bibitem[\protect\citeauthoryear{{West}, {Hawley}, {Bochanski}, {Covey},
  {Reid}, {Dhital}, {Hilton}  \& {Masuda}}{{West} et~al.}{2008}]{West08}
{West} A.~A.,  {Hawley} S.~L.,  {Bochanski} J.~J.,  {Covey} K.~R.,  {Reid}
  I.~N.,  {Dhital} S.,  {Hilton} E.~J.,   {Masuda} M.,  2008, \mn@doi [\aj]
  {10.1088/0004-6256/135/3/785}, \href
  {http://adsabs.harvard.edu/abs/2008AJ....135..785W} {135, 785}

\bibitem[\protect\citeauthoryear{{Wright} et~al.,}{{Wright}
  et~al.}{2010a}]{Wright10}
{Wright} E.~L.,  et~al., 2010a, \mn@doi [\aj] {10.1088/0004-6256/140/6/1868},
  \href {http://adsabs.harvard.edu/abs/2010AJ....140.1868W} {140, 1868}

\bibitem[\protect\citeauthoryear{{Wright} et~al.,}{{Wright}
  et~al.}{2010b}]{WISE2010}
{Wright} E.~L.,  et~al., 2010b, \mn@doi [\aj] {10.1088/0004-6256/140/6/1868},
  \href {http://adsabs.harvard.edu/abs/2010AJ....140.1868W} {140, 1868}

\bibitem[\protect\citeauthoryear{{York} et~al.,}{{York} et~al.}{2000}]{York00}
{York} D.~G.,  et~al., 2000, \mn@doi [\aj] {10.1086/301513}, \href
  {http://adsabs.harvard.edu/abs/2000AJ....120.1579Y} {120, 1579}

\makeatother
\end{thebibliography}

\clearpage
\onecolumn
\appendix

\section{Configuration files}\label{app.files}

\centering
\begin{longtable}{lr}
\caption{Configuration parameters for {\sc SExtractor}.}\\
\label{tab.def_sex}\\
\hline
\noalign{\smallskip}
\endfirsthead
\caption{Configuration parameters for {\sc SExtractor} (continued).}\\
\hline
\noalign{\smallskip}
\endhead
\multicolumn{2}{c}{Catalog}	\\
\noalign{\smallskip}
\hline 
 \noalign{\smallskip}
CATALOG\_NAME   &  test.cat      \\
CATALOG\_TYPE    &FITS\_LDAC    \\
PARAMETERS\_NAME &default.param \\
\noalign{\smallskip}
\hline
\noalign{\smallskip}
\multicolumn{2}{c}{Extraction}  \\
\noalign{\smallskip}
\hline
\noalign{\smallskip}
DETECT\_TYPE     &CCD           \\
DETECT\_MINAREA  &3             \\
DETECT\_MAXAREA  &0             \\
THRESH\_TYPE     &RELATIVE      \\
DETECT\_THRESH   &1.5          \\
ANALYSIS\_THRESH &1.5          \\
 
FILTER          &Y             \\
FILTER\_NAME     &gauss\_2.5\_5x5.conv  \\
FILTER\_THRESH   &      --       \\
 
DEBLEND\_NTHRESH &32           \\
DEBLEND\_MINCONT &0.0005       \\
 
CLEAN           &Y            \\
CLEAN\_PARAM     &1.0          \\
 
MASK\_TYPE       &CORRECT      \\

\noalign{\smallskip}
\hline
\noalign{\smallskip}
\multicolumn{2}{c}{WEIGHTing}   \\
\noalign{\smallskip}
\hline
\noalign{\smallskip}
WEIGHT\_TYPE     &MAP\_WEIGHT    \\
WEIGHT\_IMAGE    &weight.fits   \\
WEIGHT\_GAIN     &Y             \\
WEIGHT\_THRESH   &      --       \\

\noalign{\smallskip}
\hline
\noalign{\smallskip}
\multicolumn{2}{c}{FLAGging}    \\
\multicolumn{2}{c}{Photometry}  \\
\noalign{\smallskip}
\hline
\noalign{\smallskip}
PHOT\_APERTURES  &20,25,30        \\
PHOT\_AUTOPARAMS &2.5, 3.5      \\
PHOT\_PETROPARAMS&2.0, 3.5      \\
PHOT\_AUTOAPERS  &0.0,0.0      \\
PHOT\_FLUXFRAC   &0.5          \\

SATUR\_LEVEL     &62000.0      \\
SATUR\_KEY       &DUMMY        \\
 
MAG\_ZEROPOINT   &0.0          \\
MAG\_GAMMA       &4.0          \\
GAIN            &0.0          \\
GAIN\_KEY        &GAIN        \\
PIXEL\_SCALE     &0.0          \\
\noalign{\smallskip}
\hline
\noalign{\smallskip}
\multicolumn{2}{c}{Star/Galaxy Separation}  \\
\noalign{\smallskip}
\hline
\noalign{\smallskip}
SEEING\_FWHM     &0.8          \\
STARNNW\_NAME    &default.nnw  \\
\noalign{\smallskip}
\hline
\noalign{\smallskip}
\multicolumn{2}{c}{Background}  \\
\noalign{\smallskip}
\hline
\noalign{\smallskip}

BACK\_TYPE       &AUTO          \\
BACK\_VALUE      &0.0           \\
BACK\_SIZE       &64           \\
BACK\_FILTERSIZE &3            \\
 
BACKPHOTO\_TYPE  &GLOBAL       \\
BACKPHOTO\_THICK &24           \\
BACK\_FILTTHRESH &0.0          \\
\noalign{\smallskip}
\hline
\noalign{\smallskip}
\multicolumn{2}{c}{Check Image} \\
\noalign{\smallskip}
\hline
\noalign{\smallskip}

CHECKIMAGE\_TYPE &NONE        \\

CHECKIMAGE\_NAME &check.fits , aper.fits  \\
 
\noalign{\smallskip}
\hline
\noalign{\smallskip}
\multicolumn{2}{c}{Memory}  \\
\noalign{\smallskip}
\hline
\noalign{\smallskip}

MEMORY\_OBJSTACK &3000         \\
MEMORY\_PIXSTACK &300000      \\
MEMORY\_BUFSIZE  &1024        \\
 
\noalign{\smallskip}
\hline
\noalign{\smallskip}
\multicolumn{2}{c}{Miscellaneous}   \\
\noalign{\smallskip}
\hline
\noalign{\smallskip}

VERBOSE\_TYPE    &NORMAL       \\
HEADER\_SUFFIX   &.head        \\
WRITE\_XML       &N            \\
XML\_NAME        &sex.xml      \\
XSL\_URL         &file:///usr/share/sextractor/sextractor.xsl                             \\
NTHREADS        &1            \\

FITS\_UNSIGNED   &N            \\
INTERP\_MAXXLAG  &16           \\
INTERP\_MAXYLAG  &16          \\
INTERP\_TYPE     &NONE       \\

\noalign{\smallskip}
\hline
\noalign{\smallskip}
\multicolumn{2}{c}{Experimental Stuff}  \\
\noalign{\smallskip}
\hline
\noalign{\smallskip}

PSF\_NAME        &27sfb\_barD\_0163\_ast-red.psf  \\
PSF\_NMAX        &1             \\
PATTERN\_TYPE    &RINGS-HARMONIC \\
SOM\_NAME        &default.som   \\
\noalign{\smallskip}
\hline
\end{longtable}

\newpage
\centering
\begin{longtable}{lr}
\caption{Configuration paramerers for {\sc PSFEx}.}\\
\label{tab.def_psfex}\\
\hline
\noalign{\smallskip}
\endfirsthead
\caption{Configuration paramerers for {\sc PSFEx} (continued).}\\
\hline
\noalign{\smallskip}
\endhead
\multicolumn{2}{c}{PSF model}	\\
 \noalign{\smallskip}
 \hline
 \noalign{\smallskip}
BASIS\_TYPE     &PIXEL\_AUTO \\
BASIS\_NUMBER   &20           \\
BASIS\_NAME     &basis.fits   \\
BASIS\_SCALE    &1.0            \\
NEWBASIS\_TYPE  &NONE           \\
NEWBASIS\_NUMBER&8             \\
PSF\_SAMPLING   &0.0           \\
PSF\_PIXELSIZE  &1.0           \\
PSF\_ACCURACY   &0.01          \\
PSF\_SIZE       &31,31         \\
PSF\_RECENTER   &Y             \\
MEF\_TYPE       &INDEPENDENT   \\
 \noalign{\smallskip}
\hline
\noalign{\smallskip}

\multicolumn{2}{c}{Point source measurements}	\\ 
\noalign{\smallskip}
\hline
\noalign{\smallskip}
 
CENTER\_KEYS    &X\_IMAGE,Y\_IMAGE \\
PHOTFLUX\_KEY   &FLUX\_APER(1)    \\
PHOTFLUXERR\_KEY&FLUXERR\_APER(1) \\
\noalign{\smallskip}
\hline
\noalign{\smallskip}
 
\multicolumn{2}{c}{PSF variability}	\\ 
\noalign{\smallskip}
\hline
\noalign{\smallskip}
 
PSFVAR\_KEYS    &X\_IMAGE,Y\_IMAGE  \\
PSFVAR\_GROUPS  &1,1             \\
PSFVAR\_DEGREES &3              \\
PSFVAR\_NSNAP   &9              \\
HIDDENMEF\_TYPE &COMMON         \\
STABILITY\_TYPE &EXPOSURE        \\
\noalign{\smallskip}
\hline
\noalign{\smallskip}
 
\multicolumn{2}{c}{Sample selection}	\\
\noalign{\smallskip}
\hline
\noalign{\smallskip}
 
SAMPLE\_AUTOSELECT &Y            \\
SAMPLEVAR\_TYPE    &SEEING       \\
SAMPLE\_FWHMRANGE  &2.0,15.0    \\
SAMPLE\_VARIABILITY&0.2        \\
SAMPLE\_MINSN      &20          \\
SAMPLE\_MAXELLIP   &0.3         \\
SAMPLE\_FLAGMASK   &0x00fe      \\
SAMPLE\_WFLAGMASK  &0x00ff      \\
SAMPLE\_IMAFLAGMASK&0x0         \\
BADPIXEL\_FILTER   &N           \\
BADPIXEL\_NMAX     &0           \\
\noalign{\smallskip}
\hline
\noalign{\smallskip}
 
\multicolumn{2}{c}{PSF homogeneisation kernel}	\\ 
\noalign{\smallskip}
\hline
\noalign{\smallskip}

HOMOBASIS\_TYPE    &NONE         \\
HOMOBASIS\_NUMBER  &10          \\
HOMOBASIS\_SCALE   &1.0          \\
HOMOPSF\_PARAMS    &2.0, 3.0     \\
HOMOKERNEL\_DIR    &   --          \\
HOMOKERNEL\_SUFFIX &.homo.fits   \\
\noalign{\smallskip}
\hline
\noalign{\smallskip}

\multicolumn{2}{c}{Check-plots}	\\ 
\noalign{\smallskip}
\hline
\noalign{\smallskip}
 
CHECKPLOT\_DEV     & PNG         \\
CHECKPLOT\_RES     & 0          \\
CHECKPLOT\_ANTIALIAS&Y          \\
CHECKPLOT\_TYPE     &NONE \\
CHECKPLOT\_NAME     &fwhm, ellipticity, counts, countfrac, chi2, resi \\%
\noalign{\smallskip}
\hline
\noalign{\smallskip}
 
\multicolumn{2}{c}{Check-Images}	\\ 
\noalign{\smallskip}
\hline
\noalign{\smallskip}
 
CHECKIMAGE\_TYPE NONE & \\
CHECKIMAGE\_NAME& chi.fits,proto.fits,samp.fits,resi.fits,snap.fits	\\
CHECKIMAGE\_CUBE N &           \\
\noalign{\smallskip}
\hline
\noalign{\smallskip}
 
\multicolumn{2}{c}{ Miscellaneous}	\\
\noalign{\smallskip}
\hline
\noalign{\smallskip}
 
PSF\_DIR         &     --          \\
PSF\_SUFFIX     & .psf           \\
VERBOSE\_TYPE   & NORMAL        \\
WRITE\_XML      & Y             \\
XML\_NAME       & psfex.xml     \\
XSL\_URL       &  file:///usr/share/psfex/psfex.xsl \\
NTHREADS       & 4             \\
 \noalign{\smallskip}
\hline
\end{longtable}

\newpage
\centering
\begin{table}
\caption{Parameters obtained from {\sc SExtractor} and {\sc PSFEx}.}
\label{tab.def_params}
\begin{tabular}{lll}
\hline\hline
\noalign{\smallskip}
NUMBER	&   Running object number   &   \\
ALPHA\_J2000	& Right Ascension of barycenter (J2000)  	&  [deg] 	\\
DELTA\_J2000	&   Declination of barycenter (J2000)	&  [deg] 	\\
XPSF\_IMAGE		&   X coordinate from PSF-fitting	& [pixel]  \\
YPSF\_IMAGE	&   Y coordinate from PSF-fitting	&  [pixel]  \\
ERRAPSF\_IMAGE	&   PSF RMS position error along major axis	&   [pixel] 	\\
ERRBPSF\_IMAGE	&   PSF RMS position error along minor axis	&   [pixel] 	\\
ERRTHETAPSF\_IMAGE	&   PSF error ellipse position angle (CCW/x)	&   [deg]	\\
XWIN\_IMAGE	&  Windowed position estimate along x 	&  [pixel] 	\\
YWIN\_IMAGE	&   Windowed position estimate along y	&   [pixel]	\\
ERRAWIN\_IMAGE	&   RMS windowed position error along major axis	&   [pixel]	\\
ERRBWIN\_IMAGE	&   RMS windowed position error along minor axis	&   [pixel]	\\
ERRTHETAWIN\_IMAGE	&   Windowed error ellipse position angle (CCW/x)	&   [deg]	\\
FLUX\_AUTO	&   Flux within a Kron-like elliptical aperture	&   [count]	\\
FLUXERR\_AUTO	&   RMS error for AUTO flux	&   [count]	\\
MAG\_AUTO	&   Kron-like elliptical aperture magnitude	&   [mag]   \\
MAGERR\_AUTO	&   RMS error for AUTO magnitude	&   [mag]	\\
FLUX\_APER	&   Flux vector within fixed circular aperture(s)	&   [count]	\\
FLUXERR\_APER	&   RMS error vector for aperture flux(es)	&   [count]	\\
MAG\_APER	&   Fixed aperture magnitude vector	&   [mag]	\\
MAGERR\_APER	&   RMS error vector for fixed aperture magnitude	&  [mag] 	\\
FLUX\_PSF	&   Flux from PSF-fitting	&   [count]	\\
FLUXERR\_PSF	&   RMS flux error for PSF-fitting	&   [count]	\\
MAG\_PSF	&   Magnitude from PSF-fitting	&   [mag]	\\
MAGERR\_PSF	&   RMS magnitude error from PSF-fitting	&   [mag]	\\
FWHM\_IMAGE	&   FWHM assuming a gaussian core	&   [pixel]	\\
ELONGATION	&   A\_IMAGE/B\_IMAGE (where A\_IMAGE is the profile RMS along the major	&	\\
			&  axis and B\_IMAGE is the profile RMS along the minor axis) 	&   	\\
ELLIPTICITY	&   1 - B\_IMAGE/A\_IMAGE(where A\_IMAGE is the profile RMS along the major	&	\\
			&  axis and B\_IMAGE is the profile RMS along the minor axis) 	&   	\\
FLUX\_RADIUS	&   Fraction-of-light radii	&   [pixel]	\\
FLAGS	&   Extraction flags	&   	\\
FLAGS\_WEIGHT	&   Weighted extraction flags	&   	\\
SPREAD\_MODEL	&   Spread parameter from model-fitting	&   	\\
SPREADERR\_MODEL	&   Spread parameter error from model-fitting	&   	\\
XMODEL\_IMAGE	&   X coordinate from model-fitting	&   [pixel]	\\
YMODEL\_IMAGE	&   Y coordinate from model-fitting	&   [pixel]	\\
ERRAMODEL\_IMAGE	&   RMS error of fitted position along major axis	& [pixel]  	\\
ERRBMODEL\_IMAGE	&   RMS error of fitted position along minor axis	&   [pixel]	\\
ERRTHETAMODEL\_IMAGE	&   Error ellipse position angle of fitted position (CCW/x)	&   [deg]	\\
FLUX\_MODEL	&   Flux from model-fitting	&   [count]	\\
FLUXERR\_MODEL	&   RMS error on model-fitting flux	&  [count] 	\\
FLAGS\_MODEL	&   Model-fitting flags	&   	\\
NITER\_MODEL	&   Number of iterations for model-fitting	&   	\\
EXT\_NUMBER	&   FITS extension number	&   	\\
\noalign{\smallskip}
\hline
\end{tabular}
\end{table}

\clearpage

\section{Catalogue description}\label{app.cats}

\clearpage

\centering
\begin{table}
\caption{Description of the parameters contained in the detection catalogue.}
\label{tab.catalogue_description}
\begin{tabular}{p{2.5cm}p{1cm}p{13cm}}
\hline \hline
	\noalign{\smallskip}
Parameter    &   Units   &   Description \\
	\noalign{\smallskip}

	\hline
RAJ2000 &   deg &   Right Ascension (J2000).   \\
DEJ2000 &   deg &   Declination (J2000).   \\
	\noalign{\smallskip}
eRA &   arcsec  & Right Ascension uncertainty ($e\_RAJ2000*\cos{\rm{DEJ2000}}$).\\
eDE &   arcsec  & Declination uncertainty.  \\
	\noalign{\smallskip}
ID\_source  & --  &   Unique source identifier. It follows an IAU-style designation of the form "GTC$_{-}$OSIRIS$_{-}$BBI$_{-}$DR1$_{-}$JHHMMSS.ss+DDMMSS.s", where "GTC$_{-}$OSIRIS$_{-}$BBI$_{-}$DR1" refers to the telescope (\gtc), the instrument (\osiris), the observing mode (Broad Band Image) and data release (DR1). The remaining string denotes the J2000 coordinates in sexagesimal format.    \\
	\noalign{\smallskip}
ID\_detection   &  -- &  Detection identifier composed by the {\it ID\_source} and "\_n", where n is an integer going from 1 to the total number of detections of the source.\\
	\noalign{\smallskip}
Image\_url   & --  &  URL access to the FITS image in which the detection has been made.  \\
	\noalign{\smallskip}
Xmag\_psf   &   mag & PSF calibrated magnitude. {\it X} stands for $g$, $r$, $i$ and $z$.    \\
eXmag\_psf  &   mag & PSF calibrated magnitude error. {\it X} stands for $g$, $r$, $i$ and $z$.  \\
	\noalign{\smallskip}
Flag\_psf   & --    &    Flag for PSF calibrated magnitude. \\
            &   &   "A" stands for PSF calibrated magnitudes within the interval of magnitudes used for the photometric calibration.    \\
            &   &   "B" stands for PSF calibrated magnitudes fainter than the faintest magnitude used for the photometric calibration.    \\
            &   &   "C" stands for PSF calibrated magnitudes brighter than the brightest magnitude used for the photometric calibration.    \\
	\noalign{\smallskip}
Xmag\_auto   &   mag & AUTO calibrated magnitude. {\it X} stands for $g$, $r$, $i$ and $z$.    \\
eXmag\_auto  &   mag & AUTO calibrated magnitude error. {\it X} stands for $g$, $r$, $i$ and $z$.  \\
	\noalign{\smallskip}
Flag\_auto   & --  &    Flag for AUTO calibrated magnitude. \\
            &   &   "A" stands for AUTO calibrated magnitudes within the interval of magnitudes used for the photometric calibration.    \\
            &   &   "B" stands for AUTO calibrated magnitudes fainter than the faintest magnitude used for the photometric calibration.    \\
            &   &   "C" stands for AUTO calibrated magnitudes brighter than the brightest magnitude used for the photometric calibration.    \\
	\noalign{\smallskip}
Xmag\_aperY\_inst   &  mag &   Instrumental aperture photometry. {\it X} stands for the four $g$, $r$, $i$, $z$ and {\it Y} stands for the three apertures defined in Section~\ref{sec:sources_extraction}.   \\
eXmag\_aperY\_inst   &  mag &   Instrumental aperture photometric errors. {\it X} stands for $g$, $r$, $i$, $z$ and {\it Y} stands for the three apertures defined in Section~\ref{sec:sources_extraction}.   \\
	\noalign{\smallskip}
Xmag\_model\_inst   &   mag &   Instrumental \verb|MODEL| photometry. {\it X} stands for the four $g$, $r$, $i$, $z$. \\
eXmag\_model\_inst   &   mag &   Instrumental \verb|MODEL| photometric errors. {\it X} stands for the four $g$, $r$, $i$, $z$. \\
	\noalign{\smallskip}
Elongation  &   --  &  Elongation of the detection defined as $A/B$.   \\ 
	\noalign{\smallskip}
Ellipticity &   --  & Ellipticity of the detection defined as $1 - B/A$.     \\
	\noalign{\smallskip}
FWHM        &   arcsec  &    Full width half maximum of the detection.   \\
	\noalign{\smallskip}
SNR         &   --- &    Signal to noise ratio.  \\
	\noalign{\smallskip}
MJD         &   d   &   Modified Julian Date of the observation.    \\
	\noalign{\smallskip}
Type       &   --  &   Indicates whether the detection is point-like ("P" or "P*") or extense ("E" or "E*") as explained in Section~\ref{subsec:morph}.    \\
	\noalign{\smallskip}
cl  &   --  &    Point-like coefficient linked to each unique source. It is defined as the ratio between number of point-like ("P" or "P*") detections and the total number of point-like and extense ("E" or "E*") classifications of the same source ($num_P/num_{P+E}$). $cl$ equal to 1 indicates that the source has been always classified as point-like, and $cl$ equal to 0 indicates that the source has been always classified as extense.   \\
	\noalign{\smallskip}
Flag\_source    &   -- &   Indicates if it is a {\it primary} (p) or {\it secondary} (s) detection, as defined in Section~\ref{subsec:resulting_cat}.   \\
	\noalign{\smallskip}

\hline
\end{tabular}
\end{table}

\newpage

\begin{table}
\centering
\caption{Description of the parameters contained in the source catalogue.}
\label{tab.primarycatalogue_description}
\begin{tabular}{p{2.5cm}p{1cm}p{13cm}}
\hline \hline
	\noalign{\smallskip}
Parameter    &   Units   &   Description \\
	\noalign{\smallskip}

	\hline
RAJ2000 &   deg &   Mean weighted Right Ascension (J2000).   \\
DEJ2000 &   deg &   Mean weighted Declination (J2000).   \\
	\noalign{\smallskip}
eRA &   arcsec  & Right Ascension uncertainty ($e\_RAJ2000*\cos{\rm{DEJ2000}}$ for sources with unique detection or the standard deviation of the weighted mean otherwise).\\
eDE &   arcsec  & Declination uncertainty ($e\_DEJ2000$ for sources with unique detection or the standard deviation of the weighted mean otherwise).  \\
	\noalign{\smallskip}
	cl  &   --  &    Point-like coefficient linked to each unique source. It is defined as the ratio between number of point-like ("P" or "P*") detections and the total number of point-like and extense ("E" or "E*") classifications of the same source ($num_P/num_{P+E}$). $cl$ equal to 1 indicates that the source has been always classified as point-like, and $cl$ equal to 0 indicates that the source has been always classified as extense.   \\
	\noalign{\smallskip}
ID\_source  & --  &   Unique source identifier. It follows an IAU-style designation of the form "GTC$_{-}$OSIRIS$_{-}$BBI$_{-}$DR1$_{-}$JHHMMSS.ss+DDMMSS.s", where "GTC$_{-}$OSIRIS$_{-}$BBI$_{-}$DR1" refers to the telescope (\gtc), the instrument (\osiris), the observing mode (Broad Band Image) and data release (DR1). The remaining string denotes the J2000 coordinates in sexagesimal format.    \\
	\noalign{\smallskip}
ID\_detection\_Xpsf   &  -- &  Detection identifier of the PSF photometry in the {\it X} band composed by the {\it ID\_source} and "\_n", where n corresponds with the n-th detection of the source in the source catalogue. {\it X} stands for $g$, $r$, $i$ and $z$.\\
	\noalign{\smallskip}
Image\_url\_Xpsf   & --  &  URL access to the FITS image in which the PSF detection has been made. {\it X} stands for $g$, $r$, $i$ and $z$. \\
	\noalign{\smallskip}
Xmag\_psf   &   mag & PSF calibrated magnitude. {\it X} stands for $g$, $r$, $i$ and $z$.    \\
eXmag\_psf  &   mag & PSF calibrated magnitude error. {\it X} stands for $g$, $r$, $i$ and $z$.  \\
	\noalign{\smallskip}
Flag\_Xpsf   & --    &    Flag for PSF calibrated magnitude. {\it X} stands for $g$, $r$, $i$ and $z$. \\
            &   &   "A" stands for PSF calibrated magnitudes within the interval of magnitudes used for the photometric calibration.    \\
            &   &   "B" stands for PSF calibrated magnitudes fainter than the faintest magnitude used for the photometric calibration.    \\
            &   &   "C" stands for PSF calibrated magnitudes brighter than the brightest magnitude used for the photometric calibration.    \\
	\noalign{\smallskip}
MJD\_Xpsf     &   d   &   Modified Julian Date of the observation of the PSF photometry. {\it X} stands for $g$, $r$, $i$ and $z$.  \\
	\noalign{\smallskip}
ID\_detection\_Xauto   &  -- &  Detection identifier of the AUTO photometry in the {\it X} band composed by the {\it ID\_source} and "\_n", where n is an integer going from 1 to the total number of detections of the source in the source catalogue. {\it X} stands for $g$, $r$, $i$ and $z$.\\
	\noalign{\smallskip}
Image\_url\_Xauto   & --  &  URL access to the FITS image in which the AUTO detection has been made. {\it X} stands for $g$, $r$, $i$ and $z$. \\
    \noalign{\smallskip}
Xmag\_auto   &   mag & AUTO calibrated magnitude. {\it X} stands for $g$, $r$, $i$ and $z$.    \\
eXmag\_auto  &   mag & AUTO calibrated magnitude error. {\it X} stands for $g$, $r$, $i$ and $z$.  \\
	\noalign{\smallskip}
Flag\_Xauto   & --  &    Flag for AUTO calibrated magnitude. {\it X} stands for $g$, $r$, $i$ and $z$. \\
            &   &   "A" stands for AUTO calibrated magnitudes within the interval of magnitudes used for the photometric calibration.    \\
            &   &   "B" stands for AUTO calibrated magnitudes fainter than the faintest magnitude used for the photometric calibration.    \\
            &   &   "C" stands for AUTO calibrated magnitudes brighter than the brightest magnitude used for the photometric calibration.    \\
	\noalign{\smallskip}
	MJD\_Xauto     &   d   &   Modified Julian Date of the observation of the AUTO photometry. {\it X} stands for $g$, $r$, $i$ and $z$.  \\
		\noalign{\smallskip}

\hline
\end{tabular}
\end{table}

\clearpage
\section{Table of cool dwarf candidates}\label{app.cdwarfs}
\clearpage

\centering
\begin{table}
\caption{\osiris and {\it Gaia} DR2 identifiers, and effective temperatures of the 49 cool dwarf candidates (``*'' identifies the secondary component of the close K+M binary system) plus the late-type binary (``**'' and ``***'' identify the primary and the secondary components, respectively) found in this work.}
\label{tab.cooldwarfs}
\begin{tabular}{llc}
\hline\hline
\noalign{\smallskip}
ID & {\it Gaia} DR2 source\_id    &   $T_{eff}$ [K] \\
\noalign{\smallskip}
\hline
  GTC\_OSIRIS\_BBI\_DR1\_J095528.33+694328.2 & 1070554378447970176 & 3000.0\\
  GTC\_OSIRIS\_BBI\_DR1\_J095533.01+694339.1 & 1070554378447969920 & 3200.0\\
  GTC\_OSIRIS\_BBI\_DR1\_J095612.91+693729.9 & 1070549220190318464 & 3100.0\\
  GTC\_OSIRIS\_BBI\_DR1\_J111433.38-302022.4 & 5404305499712757888 & 3100.0\\
  GTC\_OSIRIS\_BBI\_DR1\_J111449.77-302006.9 & 5404305121755637632 & 3400.0\\
  GTC\_OSIRIS\_BBI\_DR1\_J152215.23+285622.0 & 1272146188971317632 & 3200.0\\
  GTC\_OSIRIS\_BBI\_DR1\_J152224.31+290123.5 & 1272152923480070144 & 3400.0\\
  GTC\_OSIRIS\_BBI\_DR1\_J154542.83-001823.1 & 4404496211953819392 & 3300.0\\
  GTC\_OSIRIS\_BBI\_DR1\_J164751.11+715102.1 & 1653530664857859456 & 3100.0\\
  GTC\_OSIRIS\_BBI\_DR1\_J164823.16+714939.3 & 1653529973366993792 & 3300.0\\
  GTC\_OSIRIS\_BBI\_DR1\_J170533.87+112345.0 & 4444960919915917440 & 3100.0\\
  GTC\_OSIRIS\_BBI\_DR1\_J181603.76+691300.4 & 2259804228271304192 & 3300.0\\
  GTC\_OSIRIS\_BBI\_DR1\_J194704.87+313659.8 & 2034508278496366208 & 3100.0\\
  GTC\_OSIRIS\_BBI\_DR1\_J203115.87+410311.3 & 2067801078262984960 & 3200.0\\
  GTC\_OSIRIS\_BBI\_DR1\_J203116.85+410028.4 & 2067797917166997248 & 3400.0\\
  GTC\_OSIRIS\_BBI\_DR1\_J203118.11+405343.5 & 2067794962230400000 & 3100.0\\
  GTC\_OSIRIS\_BBI\_DR1\_J203146.14+412246.4 & 2067829356329503616 & 2900.0\\
  GTC\_OSIRIS\_BBI\_DR1\_J203146.23+411437.0** & 2067827608276334848 &\\
  GTC\_OSIRIS\_BBI\_DR1\_J203146.19+411437.6*** &   2067827608279535744 &   \\
  GTC\_OSIRIS\_BBI\_DR1\_J203151.94+410902.4 & 2067822484380101888 & 3200.0\\
  GTC\_OSIRIS\_BBI\_DR1\_J203152.80+412429.0 & 2067829493768462464 & 3400.0\\
  GTC\_OSIRIS\_BBI\_DR1\_J203159.58+412624.2 & 2067835511019676544 & 3700.0\\
  GTC\_OSIRIS\_BBI\_DR1\_J203204.11+410327.8 & 2067773968429422976 & 3200.0\\
  GTC\_OSIRIS\_BBI\_DR1\_J203204.73+411135.0 & 2067824064928106880 & 3300.0\\
  GTC\_OSIRIS\_BBI\_DR1\_J203213.30+412156.6 & 2067834338490023552 & 3600.0\\
  GTC\_OSIRIS\_BBI\_DR1\_J203218.88+412303.6 & 2067834166691344896 & 3500.0\\
  GTC\_OSIRIS\_BBI\_DR1\_J203228.12+410853.8 & 2067776476690427008 & 3800.0\\
  GTC\_OSIRIS\_BBI\_DR1\_J203229.46+410025.6 & 2067773040716423680 & 3300.0\\
  GTC\_OSIRIS\_BBI\_DR1\_J203240.40+410329.6 & 2067774689983932032 & 3500.0\\
  GTC\_OSIRIS\_BBI\_DR1\_J203255.73+411428.9 & 2067782901961590528 & 3200.0\\
  GTC\_OSIRIS\_BBI\_DR1\_J203302.89+412236.0 & 2067831967669718144 & 3500.0\\
  GTC\_OSIRIS\_BBI\_DR1\_J203307.46+410420.9 & 2067777855378562432 & 2400.0\\
  GTC\_OSIRIS\_BBI\_DR1\_J203317.38+411906.9 & 2067785002201372032 & 3200.0\\
  GTC\_OSIRIS\_BBI\_DR1\_J203323.48+412040.5 & 2067785135346315136 & 3000.0\\
  GTC\_OSIRIS\_BBI\_DR1\_J203324.41+410751.7* & 2067777782367558784 & 3300.0\\
  GTC\_OSIRIS\_BBI\_DR1\_J203341.70+405844.3 & 2067763802245107584 & 3300.0\\
  GTC\_OSIRIS\_BBI\_DR1\_J203344.35+412131.5 & 2067878078436914560 & 2800.0\\
  GTC\_OSIRIS\_BBI\_DR1\_J203353.39+410013.9 & 2067764244624468736 & 3100.0\\
  GTC\_OSIRIS\_BBI\_DR1\_J203402.48+413937.6 & 2067935291698824064 & 3200.0\\
  GTC\_OSIRIS\_BBI\_DR1\_J203403.33+410246.1 & 2067765893892779008 & 3300.0\\
  GTC\_OSIRIS\_BBI\_DR1\_J203403.60+412450.4 & 2067877150723292288 & 3200.0\\
  GTC\_OSIRIS\_BBI\_DR1\_J203404.41+405530.4 & 2067760087096005760 & 3200.0\\
  GTC\_OSIRIS\_BBI\_DR1\_J203408.50+411954.7 & 2067874543679715200 & 3000.0\\
  GTC\_OSIRIS\_BBI\_DR1\_J203415.44+405716.3 & 2067761637577789312 & 3000.0\\
  GTC\_OSIRIS\_BBI\_DR1\_J203416.90+411944.7 & 2067875776335180800 & 3000.0\\
  GTC\_OSIRIS\_BBI\_DR1\_J203425.02+405820.6 & 2067761598925476352 & 3600.0\\
  GTC\_OSIRIS\_BBI\_DR1\_J203430.63+411138.3 & 2067860967287122816 & 3100.0\\
  GTC\_OSIRIS\_BBI\_DR1\_J203434.94+412553.5 & 2067876708344798336 & 3600.0\\
  GTC\_OSIRIS\_BBI\_DR1\_J203446.65+405656.5 & 2064759279345218304 & 3300.0\\
  GTC\_OSIRIS\_BBI\_DR1\_J203509.14+415510.5 & 2067916801861913984 & 3200.0\\
  GTC\_OSIRIS\_BBI\_DR1\_J203515.96+410601.1 & 2064855486612035584 & 3200.0\\
\hline
\end{tabular}
\end{table}

\clearpage

\label{lastpage}
\end{document}